\documentclass[11pt]{article}

\usepackage{amssymb}
\usepackage{amsmath}
\usepackage{epsfig}
\usepackage{subfigure}
\usepackage{multirow}

\begin{document}

\title{Higher Powers in Gravitation}
\author{{Timothy Clifton\thanks{
e-mail: TClifton@astro.ox.ac.uk}}\\
{\small {\textit{Department of Astrophysics,}}}\\
{\small {\textit{University of Oxford, Oxford OX1 3RH, UK}}}}
\date{{\normalsize July 30, 2008}}
\maketitle

\begin{abstract}

We consider the Friedmann-Robertson-Walker cosmologies of theories of gravity that generalise the
Einstein-Hilbert action by replacing the Ricci
scalar, $R$, with some function, $f(R)$.  The general asymptotic
behaviour of these cosmologies is found, at both early and late
times, and the effects of adding higher and lower powers of $R$ to the
Einstein-Hilbert action is investigated.  The assumption that the
highest powers of $R$ should dominate the Universe's early history,
and that the lowest powers should dominate its future is
found to be inaccurate.  The behaviour of the general solution is
complicated, and while it can be the case that single powers
of $R$ dominate the dynamics at late times, it can be either the higher
or lower powers that do so.  It is also shown
that it is often the lowest powers of $R$ that dominate at early
times, when approach to a bounce or a Tolman solution are generic
possibilities.  Various examples are considered, and both vacuum and
perfect fluid solutions investigated.

\end{abstract}

\section{Introduction}

We study here the dynamics of Friedmann-Robertson-Walker (FRW)
universes in $f(R)$ theories of gravity.  These theories are derived
from generalisations of the usual Einstein-Hilbert Lagrangian of
General Relativity (GR), such that 
\begin{equation}
\label{gravL}
\mathcal{L} = f(R),
\end{equation}
and have been considered extensively in the
literature (see e.g. \cite{early1,early2,early3,early4}).  Specification of the function $f(R)$ defines the theory,
and GR can be seen to be the special case $f =R$.  Such theories have
drawn considerable interest as they found success in early attempts to
create a perturbatively re-normalisable quantum field theory of gravity
\cite{qgrav}, as well as turning up more recently in the effective
actions of string theory \cite{string1,string2}.  In cosmology these theories
have been used extensively in attempts to explain the late-time accelerating
expansion of the Universe \cite{accel1,accel2}, cosmological inflation
\cite{infl1,infl2,infl3} and the nature of the initial singularity
\cite{singul1,singul2,singul3}.
For a recent review see \cite{review}.

In considering generalised $f(R)$ theories of gravity
it is often implicitly assumed that at late times in the evolution of
the Universe it should be the lowest powers of $R$ that dominate the
gravitational Lagrangian.  That is, at late times we should have $R \rightarrow 0$,
and the Universe should behave as if it were governed by a gravitational
Lagrangian of the form
\begin{equation}
\label{low}
\mathcal{L}_0 = \lim_{R \to 0} f(R),
\end{equation}
which is often presumed to correspond to the Einstein-Hilbert Lagrangian,
although other limits have been considered in attempts to address the
apparent late-time acceleration of the Universe \cite{accel1,accel2}.  Conversely, the
introduction of higher powers of $R$ into the gravitational Lagrangian
has often been assumed to mean that at early times the Universe should
behave as if governed by the vacuum dynamics of a Lagrangian
\begin{equation}
\label{high}
\mathcal{L}_{\infty} = \lim_{R \to \infty} f(R).
\end{equation}
The picture is then one of a universe that starts off at high $R$,
dominated by a Lagrangian of the form (\ref{high}), and that
subsequently expands until $R$ becomes small and the gravitational
dynamics are dominated by a Lagrangian of the form (\ref{low}).
It is the purpose of this paper to determine the veracity of such
assumptions.  This is achieved by studying the dynamical evolution
of FRW universes, governed by theories with general $f(R)$.  The asymptotic behaviour of the
general solution to the Friedmann equations is then investigated, and used to
evaluate the extent to which the afore mentioned behaviour may be considered generic.

The general solutions of FRW cosmologies governed by $f(R)$ theories
of gravity have been studied previously by a number of authors, in a
number of different contexts.  Much of this work has made use of
the dynamical systems approach, which has been used to study specific
classes of $f(R)$ in isotropic
cosmologies in \cite{Power, FRW1,FRW2}, and anisotropic cosmologies in
\cite{Bianchi1,Bianchi2,Bianchi3}.  Exact analytic expressions have been found for
the general FRW solutions of some $f(R)$ theories in
\cite{exact}, and the dynamical systems approach applied to general $f(R)$ has been considered in
\cite{gen}\footnote{See \cite{crit} for a criticism of this work.  The
present study is free from the defects pointed out in \cite{crit}.}.
For studies of spherically symmetric and weak field solutions see
\cite{review,Power,weak,ppn}, and references therein.
The approach used here is a generalisation of the analysis performed
in \cite{Power}, where theories of the form $f \propto R^n$ were
considered.

We find here that the late-time attractors of FRW cosmologies for
general $f(R)$ have various different forms, and that the asymptote
toward which the general solution is attracted depends upon the initial
conditions.  Some of these solutions correspond to the
lowest powers of $f(R)$ dominating at late-times, and others to the
highest powers.  Expanding universes with powers of $R$ lower than $R^2$
dominating their dynamics generically appear to exhibit the former
behaviour, while universes with powers of $R$ greater than $R^2$ dominating
appear to generically exhibit the latter.  Expanding universes
with higher powers of $R$ dominating their early evolution therefore appear
unlikely to evolve to a state where the Einstein-Hilbert term
dominates.

We also find that there usually exist multiple early-time attractors for the general
solution.  These can take on different forms, but generically it appears
that they either evolve as $a \sim t^{\frac{1}{2}}$, toward a big bang
singularity in the past, or that they approach a point
of inflexion, where the scale factor is constant.
This is in good agreement with the analytic general solutions for $f
\propto R^n$ found in \cite{exact}.  These results do not mean that a period
of inflation cannot occur (as indeed it appears to if $f \sim R^2$),
but it does mean that the general solution does not generally start
off inflating  (even if $f \sim R^2$).  It also means that the picture of the highest powers
of $R$ dominating the earliest stages of the Universe's evolution may
not be an accurate one.

We begin in section 2 by giving the FRW field equations for $f(R)$
theories, together with some simple power-law particular solutions
that will later appear as asymptotes of the general solution.  In
section 3 we use a dynamical systems approach to determine the form of
the general solution for vacuum cosmologies.  The phase space of the
general solution is two dimensional, and the location and stability of
critical points in this space is determined.  In section 4 we perform a
similar analysis for the case of perfect
fluid cosmologies.  The phase space of the general solution is now three
dimensional, and the location and stability of critical points is
again determined.  In section 5 we consider the effect of adding
higher and lower powers of $R$ to the Einstein-Hilbert action.
Section 6 provides a discussion of the results, and the appendix gives
some special cases that are of particular interest.

\section{f(R) Cosmology}

\subsection{Field Equations}

Replacing the Ricci scalar, $R$, in the Einstein-Hilbert action by a more
general function, $f(R)$, gives the Lagrangian density
\begin{equation}
\label{action}
\mathcal{L}=f(R) + \mathcal{L}_m,
\end{equation}
where $\mathcal{L}_m$ is the Lagrangian density of matter fields.
Variation of the corresponding action, with respect to the metric,
gives the field equations
\begin{equation}
\label{field}
f_R R_{ab}-\frac{1}{2} f g_{ab}+ {f_{R;}}^{cd} (g_{ab}
g_{cd}-g_{ac} g_{bd}) = 8 \pi T_{ab},
\end{equation}
where $f_R \equiv \delta f/\delta R$, and $T_{ab}$ is the energy-momentum tensor of matter fields, defined
in the usual way.  The Lagrangian formulation of the theory
guarantees the conservation equations ${T^{ab}}_{;b}=0$.

Substituting the FRW metric into the field
equations, (\ref{field}), gives the analogue of the Friedmann equations
\begin{align}
\label{Friedmann1}
-3 \frac{\ddot{a}}{a} f_R+\frac{1}{2} f+3 \frac{\dot{a}}{a} \dot{f}_R
 &= 8 \pi \rho\\
\label{Friedmann2}
-f_R R+2 f+3 \left( \ddot{f}_R+3 \frac{\dot{a}}{a} \dot{f}_R \right)
 &= 8 \pi (4-3 \gamma) \rho,
\end{align}
where $a=a(t)$ is the scale factor, and here we have assumed the matter fields are well described by a
perfect fluid, with barotropic equation of state $p=(\gamma-1)\rho$.
The conservation equations are then, as usual,
\begin{equation}
\label{conservation}
\dot{\rho}+3 \gamma \frac{\dot{a}}{a} \rho =0,
\end{equation}
and the Ricci scalar is
\begin{equation}
\label{Ricci}
R=\frac{1}{6} \left(
\frac{\ddot{a}}{a}+\frac{\dot{a}^2}{a^2}+\frac{k}{a^2} \right),
\end{equation}
where $k$ is the constant curvature of homogeneous spatial
3-surfaces.  Henceforth, unless explicitly stated otherwise, we will
be considering only the spatially flat class of models, where $k=0$.
Spatially curved models will be investigated elsewhere.

\subsection{Particular Solutions}

The field equations (\ref{Friedmann1})-(\ref{Ricci}) have a number of
particular solutions that are of interest.  These solutions often act as
attractors, toward which the general solutions can asymptote.  Such
behaviour will be shown in subsequent sections.

First let us consider the `vacuum dominated' solution
\begin{flalign}
\label{vacuum}
\hspace{150pt}
a \sim (t-t_0)^{\frac{B_0 (1+B_0)}{1-3 B_0+2 B_0^2}}
&\qquad \qquad &\text{[vacuum dominated solution]}
\end{flalign}
where $t_0$ is a constant, and $B_0$ is defined as
\begin{equation}
B_0 \equiv 1-\frac{f}{R f_R} \rightarrow \text{constant}.
\end{equation}
That $B_0 \rightarrow$constant is a requirement on the theory in
order for the solution to exist.  Equation (\ref{vacuum}) represents a cosmology
in which the evolution of the scale factor is dominated by the
dynamics of the Ricci curvature itself.  It is this type of evolution
that is often invoked to account for the late-time acceleration of the
Universe, or its inflation at early times.  In the limit that GR
is approached, and $B_0 \rightarrow 0$, this solution
reduces to Minkowski space.

Another solution of interest is the `matter dominated' solution
\begin{flalign}
\label{matter}
\hspace{150pt}
a \sim (t-t_0)^{\frac{2}{3 \gamma (1-B_0)}}.
&\qquad \qquad &\text{[matter dominated solution]}
\end{flalign}
Again, this solution exists if $B_0 \rightarrow$constant (not
necessarily the same constant as in (\ref{vacuum})).  This solution,
if it exists, is dependent on the matter content of the space-time, as
can be seen from the explicit dependence on $\gamma$.  As $B_0 \rightarrow 0$, and GR is
approached, this solution reduces to the usual spatially flat, perfect
fluid dominated Friedmann solution.

The last particular solution of interest is the
Tolman solution
\begin{flalign}
\label{radiation}
\hspace{150pt}
a \sim (t-t_0)^{\frac{1}{2}}.
&\qquad \qquad &\text{[Tolman solution]}
\end{flalign}
This solution is independent of both the matter content of the
Universe, and of the form of $f(R)$.  We call it the Tolman solution
as it is identical to the radiation
dominated Friedmann solution of GR.  However, this solution does not
necessarily require the presence of either radiation ($\gamma=4/3$), or of an Einstein-Hilbert term
in the action.

We will now proceed to investigate the form of the general solutions
to equations (\ref{Friedmann1})-(\ref{Ricci}), first for the vacuum
case ($\rho=0$), and then for the perfect fluid case ($\rho \neq 0$).
The particular solutions above will often appear as the asymptotic
form of the general solution, as $t \rightarrow \pm \infty$ or $t_0=$constant.

\section{Vacuum Cosmologies}

\subsection{The Dynamical System}

In the case of vacuum cosmologies we have $\rho=0$.  The field equations
(\ref{Friedmann1})-(\ref{Ricci}) can now be transformed into a system
of first-order differential equations, by defining the time coordinate $d\tau \equiv \sqrt{\vert R
\vert} dt$, and the new variables $x \equiv a^{\prime}/a$, $y
\equiv f_R^{\prime}/f_R$ and $z \equiv \ln \vert R \vert$.  The
equations (\ref{Friedmann1})-(\ref{Ricci}) then become
\begin{align}
\label{vac1}
x^{\prime} &= \frac{Q}{6}-2 x^2-A x y\\
\label{vac2}
y^{\prime} &= 4 x^2 +x y -y^2-\frac{Q}{3}-A y^2\\
\label{vac3}
z^{\prime} &= 2 A y,
\end{align}
with the constraint
\begin{equation}
\label{vaccon}
x^2+x y-\frac{1}{6} BQ=0,
\end{equation}
where primes denote differentiation with respect to $\tau$, and
$Q\equiv$sign$(R)$.    These definitions ensure $\tau$ is always real, and
monotonically increasing in $t$.  The two new functionals $A=A(z)$ and
$B=B(z)$ are defined by $A \equiv f_R/2 R f_{RR}$ and $B \equiv 1-f/R
f_R$.  Specifying $f(R)$ gives $A(z)$ and $B(z)$, and the system of equations
(\ref{vac1})-(\ref{vaccon}) is closed.  However, even before
specifying $f(R)$ it is possible to determine some generic features
of the system above, and hence of vacuum $f(R)$ FRW cosmologies in
general.  The form of equations (\ref{vac1})-(\ref{vaccon}) allow us to
treat the problem as a dynamical system, in which the general solutions
for $f(R)$ vacuum FRW cosmologies are given as trajectories in the phase
space ($x$,$y$,$z$).  Such an analysis will allow insight into the behaviour of these models.

First we note that the surface $R=0$ is an invariant sub-manifold of
the ($x$,$y$,$z$) phase space.  From (\ref{vaccon}) it can be seen
that $R=0$ corresponds to $x^2+x y=0$.  Equations
(\ref{vac1})-(\ref{vaccon}) then give
\begin{equation}
(x^2+xy)^{\prime} = -y (x^2+xy) (1+2A-1/B),
\end{equation}
so there are no trajectories in ($x$, $y$,$z$) that allow $R$
to change sign along them.  This justifies our choice of $z$.

It is now convenient to perform a transformation from the infinite
plane spanned by ($x$, $y$), to a finite closed space.  This can be
achieved by the re-definitions $x\equiv r \cos \theta$ and $y \equiv r \sin \theta$,
which give
\begin{align}
\label{theta}
\theta^{\prime} &= -Q \left[ 2(1-2 B) \cos \theta +(1+B) \sin \theta \right]
\sqrt{\frac{1+ \cos (2 \theta) + \sin (2\theta )}{12 BQ}}\\
z^{\prime} &= 2 A P \sqrt{\frac{B Q}{6 \cot \theta (1+\cot \theta)}},
\label{R}
\end{align}
where the constraint (\ref{vaccon}) has been used, $P$ is the sign of $y$, and $\theta$ runs
from $0$ to $2 \pi$.  It can now be seen that, independent of the value
of $B$, there exist stationary points in ($x$,$y$) at
\begin{equation}
\theta_1=-\pi/2, \qquad \theta_2=\pi/2, \qquad \theta_3=-\pi/4 \qquad \text{and}
\qquad \theta_4=3\pi/4.
\end{equation}
In the limit $B \rightarrow B_0=$constant or $\pm \infty$, if this
exists, there are two further stationary points in the ($x$,$y$) plane at
\begin{equation}
\label{56}
\tan \theta_{5,6} = \frac{2 (2 B_0-1)}{(B_0+1)}.
\end{equation}
The value of $B_0$, and the existence of these last two points, depends on the form of $f(R)$, and
can be straight-forwardly deduced once this function is specified.
Various cases will be considered in later sections.  First we will investigate the
behaviour of the scale factor at the points $1-6$, identified above.

From the definitions of $x$ and $y$ we have, at stationary $\theta_i$, that
\begin{equation}
\frac{f_R^{\prime}}{f_R} =\tan \theta_i \frac{a^{\prime}}{a} ,
\end{equation}
which can be integrated to obtain $f_R\propto a^{\tan \theta_i}$.
Eliminating $f$ in equations (\ref{Friedmann1}) and
(\ref{Friedmann2}), and substituting for $f_R$, then gives
\begin{equation}
\frac{\ddot{a}_i}{a_i} = \frac{(2+2 \tan \theta_i -\tan^2 \theta_i)}{(2+\tan
  \theta_i)} \frac{\dot{a}^2_i}{a^2_i},
\end{equation}
which, for $\tan \theta_i \neq 0$ or $1$, integrates to
\begin{equation}
\label{a}
a_i = a_0 \vert t-t_0 \vert^{\frac{(2+\tan \theta_i)}{\tan \theta_i (\tan \theta_i -1)}},
\end{equation}
where $a_0$ and $t_0$ are constants.  Substituting $\theta_i$ from
points $1-6$ then gives the form of the scale
factor as these points are approached.  For the special cases $\tan \theta_i = 0$ or $1$ we instead
have
\begin{equation}
\label{exp}
a_i = a_0 e^{c (t-t_0)},
\end{equation}
where $c$ is a constant.  For points $5$ and $6$
this corresponds to $B_0=1/2$ or $1$.

The stationary points $1$ and $2$ can now be seen to correspond to
$a^{\prime}/a=0$, and points $3$ and $4$ to the Tolman solution,
(\ref{radiation}).  For $B \neq 1/2$ or $1$ we find points $5$ and $6$
correspond to the vacuum dominated solution, (\ref{vacuum}).  These results are
summarised in Table \ref{table1}, where symbols refer to those
used in Figure \ref{vacplot}.
\begin{table}
\begin{center}
\begin{tabular}{|c|c|c|}
\hline
\textbf{Point} & \textbf{a(t)} & \textbf{Symbol}\\ \hline
1, 2 & $a^{\prime}/a=0$ & \text{Circle}\\ 
3, 4 & Tolman solution, (\ref{radiation}) & \text{Square}\\ 
5, 6 & Vacuum dominated solution, (\ref{vacuum}) & \text{Triangle}\\ \hline 
\end{tabular}
\end{center}
\caption{The evolution of the scale factor, $a(t)$, at the critical points
  $1$-$6$ in vacuum $f(R)$ cosmologies.  The
  listed symbols correspond to those used in Figure \ref{vacplot}.}
\label{table1}
\end{table}

\subsection{Stability Properties}

Having determined the location of the stationary points in the
($x$,$y$) plane, and the scale factors to which they correspond, we
will now investigate their stability properties.  This will allow us
to determine which points are stable, and can act as asymptotic
attractors of the general solution, and which are unstable, and can
act as repellors (i.e. attractors in the asymptotic past, if time is
run backwards).

The stability of points $1$ to $6$ can be determined
as follows.  Begin by perturbing $\theta$ as
\begin{equation}
\theta \rightarrow \theta_i + \delta \theta_i,
\end{equation}
where $\delta \theta_i$ is small.  For points at finite $z=z_i$ also perturb $z$, $A$ and $B$ as
\begin{align}
z &\rightarrow z_i +\delta z_i\\
A &\rightarrow A(z_i) + \left(\frac{dA}{dz} \right)_{z=z_i} \delta z_i\\
B &\rightarrow B(z_i) + \left(\frac{dB}{dz} \right)_{z=z_i} \delta z_i.
\end{align}
For points at infinite $z$ we need only to perturb
$\theta$, and to check the sign of $z^{\prime}$, in order to determine
stability.

Points $1$ and $2$, to linear order in perturbations, then give
\begin{align}
\delta \theta_{1,2}^{\prime} &\rightarrow
\mp \frac{(1+B_{1,2})}{2 B_{1,2}} \sqrt{\frac{Q B_{1,2}}{3 (1+ \cos (2 \theta_{1,2})
    +\sin (2 \theta_{1,2}))}} \delta \theta_{1,2}\\ &\sim \text{sign}
\left\{ \mp \frac{(1+B_{1,2})}{
  B_{1,2}} \right\} \delta \theta_{1,2}
\end{align}
where the upper sign is for point $1$, and the lower sign for
point $2$.  This expression is independent of any $\delta z$, and so is
valid for all points at finite or infinite $z$.

For points $3$ and $4$ we have
\begin{align}
\delta \theta_{3,4}^{\prime} &\rightarrow
\pm \frac{(5B_{3,4}-1)}{4 B_{3,4}} \sqrt{\frac{2 Q B_{3,4}}{3 (1+ \cos (2 \theta_{3,4})
    +\sin (2 \theta_{3,4}))}} \delta \theta_{3,4}\\ &\sim \text{sign}
\left\{ \pm \frac{(5B_{3,4}-1)}{B_{3,4}} \right\} \delta \theta_{3,4},
\end{align}
where the upper sign is for point $3$, and the lower for point $4$.
Again, this is independent of any $\delta z$.
Points $1$ to $4$ are therefore stable attractors if $\delta
\theta^{\prime}/\delta \theta <0$, and unstable repellors if $\delta
\theta^{\prime}/\delta \theta >0$.

Now consider the stability of points $5$ and $6$.  In
this case a knowledge of the behaviour of $z$ is necessary to
determine stability in the ($x$,$y$) plane.  For $B_0 \neq 1/2$ or $1$ the Ricci scalar
corresponding to the vacuum solution (\ref{vacuum}) is given by
\begin{align}
R_{5,6} &=\frac{B_0 (1+B_0)(5 B_0-1)}{6 (1-3B_0+2B_0^2)^2 (t-t_0)^2}\\
&\propto \exp \left\{ \mp 2\sqrt{\frac{6 (1-3
    B_0+2B_0^2)^2}{\vert B_0(1+B_0)(5B_0-1) \vert}}
(\tau-\tau_0)\right\},
\label{R56}
\end{align}
where the upper branch in the final line is for $t>t_0$, and the lower
branch for $t<t_0$.  As $\tau \rightarrow \pm \infty$, we then have
that $R_{5,6} \rightarrow \pm \infty$ or $0$.  These points therefore exist at finite $\theta$ and infinite
$z$.  As such we can check their stability by perturbing $\theta$, as
before, and checking the sign of $z^{\prime}$ as $\tau \rightarrow
\pm \infty$.  From (\ref{R56}) it can be immediately seen that for
$t>t_0$ we have $z^{\prime} <0$, so when $\delta \theta^{\prime}/\delta
\theta <0$ this is a stable point in ($x$,$y$,$z$), and corresponds 
to $R\rightarrow 0$ as $\tau \rightarrow \infty$.  Similarly, for
$t>t_0$ it can be seen that when $\delta \theta^{\prime}/\delta \theta
>0$ this is an
unstable point, corresponding to $R \rightarrow \pm
\infty$ as $\tau \rightarrow -\infty$.  The condition for the existence of points $5$ and
$6$, when $t>t_0$, can now be seen to be $B\rightarrow B_0$
as $z \rightarrow \mp \infty$, for stable or unstable points,
respectively.  Similar behaviour can be seen to be true when $t<t_0$,
with the opposite limits taken for $z$ in each case.
Substituting $\theta \rightarrow \theta_{5,6} +\delta \theta_{5,6}$ into (\ref{theta}) now
gives
\begin{align}
\delta \theta_{5,6}^{\prime} &\rightarrow
\mp \frac{(5B_0-1)}{B_0 \sqrt{1+\frac{4 (1-2 B_0)^2}{(1+B_0)^2}}} \sqrt{\frac{Q B_0}{3 (1+ \cos (2 \theta_0)
    +\sin (2 \theta_0))}} \delta \theta_{5,6}\\ &\sim \text{sign} \left\{ \mp \frac{(5B_0-1)}{
  B_0} \right\} \delta \theta_{5,6} .
\end{align}
The upper sign here corresponds to $\theta_5 \in (-\pi/2,\pi/2)$, and
the lower sign to $\theta_6 \in (\pi/2,3 \pi/2)$.  For the special
cases $B_0=1/2$ or $1$ the asymptotic value of $z$ is a finite
constant.  In these cases the stability analysis must
include a perturbation to $z$, and the corresponding perturbations
this induces in $A(z)$ and $B(z)$.  We perform this analysis in Appendix
\ref{appendix}.  For these cases we find that it is possible for
points $5$ or $6$ to initially act as attractors,
while later acting as saddle points.  We call this behaviour
`semi-stable', and describe it more fully in the appendix.

The stability properties of the critical points $1-6$ are
summarised in Table \ref{table2}.  Here `A' stands for (stable) Attractor,
and `R' for (unstable) Repellor.
\begin{table}
\begin{center}
\begin{tabular}{|c|c|c|c|c|}
\hline
\multirow{2}{*}{\textbf{Point}} &
\multicolumn{4}{|c|}{\textbf{Stability}}\\[3pt] \cline{2-5}
& $\mathbf{B<-1}$ & $\mathbf{-1<B<0}$ & $\mathbf{0<B<1/5}$ & $\mathbf{1/5<B}$ \\ \hline
1 & A & R & A & A \\
2 & R & A & R & R  \\
3 & R & R & A & R \\
4 & A & A & R & A  \\
5 & A & A & R & A  \\
6 & R & R & A & R \\ \hline
\end{tabular}
\end{center}
\caption{The stability properties of the critical points
  $1$-$6$ in vacuum $f(R)$ cosmologies. `A' denotes a stable
  Attractor, and `R' denotes an unstable Repellor.}
\label{table2}
\end{table}

\subsection{The General Solution}

\begin{figure}[htb]
\begin{center}
\subfigure[The ($x$,$y$) plane when $B<-1$ or $B>1/5$.  In this case
  the triangles, corresponding to the critical points (\ref{crit}),
  are located in the intervals ($-\pi/4$,$\pi/2$) and ($3
  \pi/4$,$-\pi/2$).  Arrows denote the direction of
  trajectories around the circle.]{\epsfig{figure=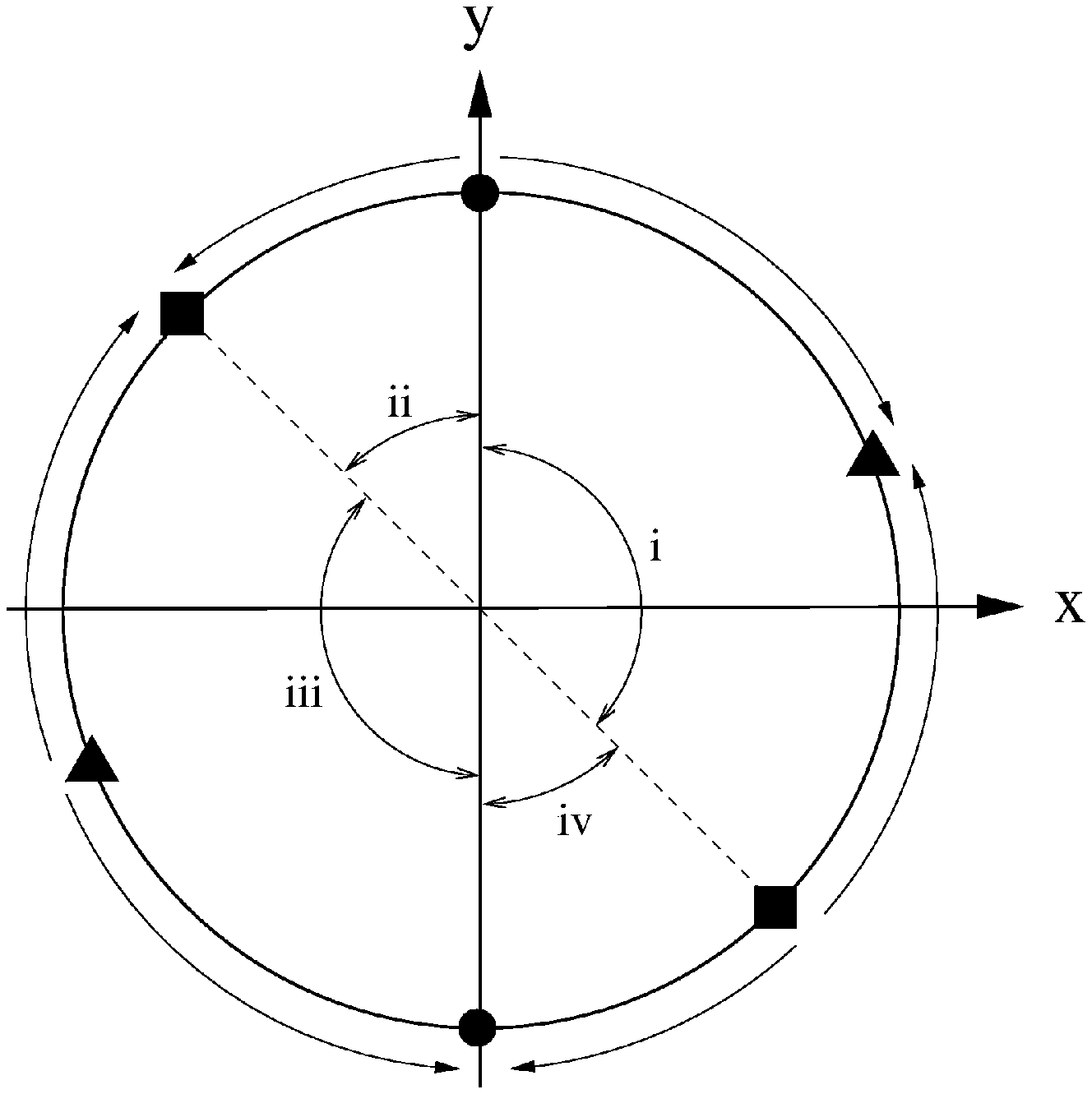,height=7cm}}
\qquad
\subfigure[The ($x$,$y$) plane when $-1<B<1/5$.  Triangles now correspond to (\ref{crit}),
  and are located in the intervals ($-\pi/2$,$-\pi/4$) and ($\pi/2$,$3
  \pi/4$).  The dashed arrows denote the direction of
  trajectories when $0<B<1/5$, and the dotted ones when $-1<B<0$.]{\epsfig{figure=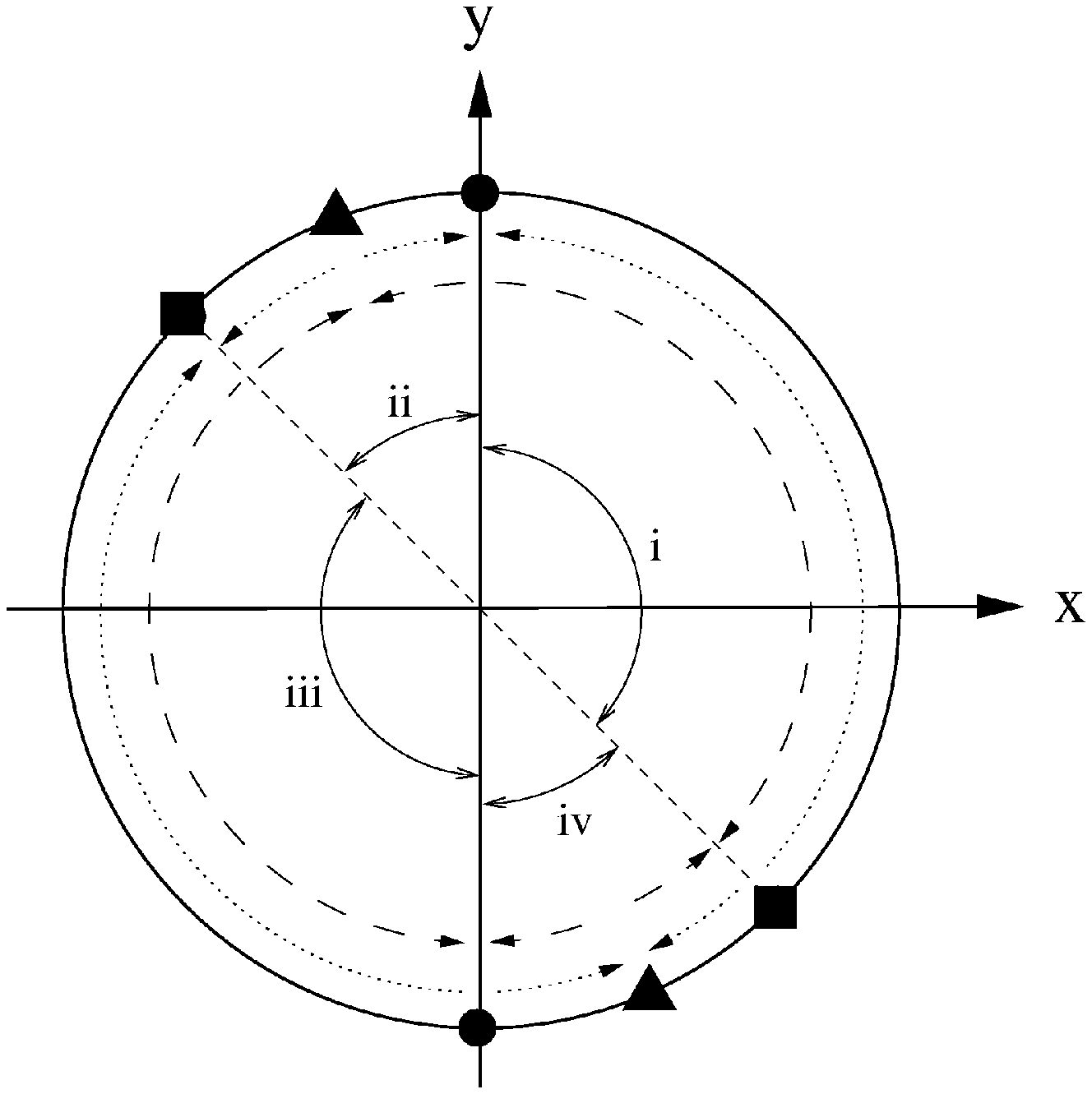,height=7cm}}
\caption{The ($x$,$y$) plane in the phase space of vacuum solutions,
  defined by equations (\ref{vac1})-(\ref{vac3}).  The arrows indicate the
  trajectories of solutions,
  confined to the circle in each plot by the constraint
  equation, (\ref{vaccon}).  Stationary points are indicated by dots,
  squares and triangles, and divide the phase space into the four
  sub-spaces $i$ to $iv$,
  between which trajectories cannot pass.  Dots correspond to
  the points $1$ and $2$, and do not move.  Squares
  correspond to points $3$ and $4$, and also do not move.  The
  Triangles correspond to the critical points defined by (\ref{crit}),
  which can move.  They approach the
  stationary points $5$ and $6$ in the limit $B\rightarrow
  B_0=$constant.}
\label{vacplot}
\end{center}
\end{figure}

We now know the location and stability properties of all stationary
points in the ($\theta$,$z$) phase space.  The possible trajectories
of solutions in this space can therefore be deduced, and are represented in the plots of Figure \ref{vacplot},
which show slices through the ($x$,$y$) plane.  Trajectories on these
diagrams are confined to the black circles, and are separated into four
sub-spaces by the circles and squares that denote
$\theta_{1,2}$ and $\theta_{3,4}$, respectively.  These four regions are
labelled $i$ to $iv$, and trajectories that
begin in one region are confined to that region for all $\tau$.  The triangles
correspond to the points
\begin{equation}
\label{crit}
\tan \theta_c = \frac{2 (2 B-1)}{(B+1)}.
\end{equation}
Other than points $1-4$, these are the only places at which
$\theta^{\prime}$ can be momentarily zero.  The triangles can move, as
$R$ evolves, and in the limit $B\rightarrow B_0$ they approach the
stationary points $5$ and $6$.  Plot \ref{vacplot}(a)
corresponds to the cases $B<-1$ and $B>1/5$, when the triangles
are in regions $i$ and $iii$. Plot \ref{vacplot}(b)
corresponds to $-1<B<1/5$, when they are in regions $ii$ and $iv$.

Let us first consider the cases depicted in Figure \ref{vacplot}(a):
$B<-1$ and $B>1/5$.  In this figure the arrows show the
direction of trajectories.  Expanding cosmologies have $x>0$, and so are
restricted to regions $i$ and $iv$.  All trajectories in region $iv$
start at point $3$ and end at point $1$.  The other expanding trajectories, in region $i$,
begin at either point $2$ or point $3$, and all end on point $5$, if that
point exists.  If point $5$ does not
exist then either $\theta_c$ is in perpetual motion in region $i$, or
it moves out of region $i$ into either region $ii$ or $iv$.  In
the former case there is then no simple late-time attractor:
trajectories are constantly evolving toward the moving $\theta_c$. In the latter case
$B$ enters the interval $-1<B<1/5$, and the trajectories are simply
attracted to either point $3$ or point $4$, as shown in Figure
\ref{vacplot}(b), and described below.  The collapsing
cosmologies with $x<0$, and described by trajectories in regions $ii$
and $iii$, can be seen to behave in a similar way, but with the
direction of time reversed, and with points
$2$ and $3$ inter-changed with points $1$ and $4$.

Now consider the cases in Figure \ref{vacplot}(b): $-1<B<1/5$.
The triangles are now found in regions $ii$ and $iv$.  Here the dashed
arrows show the direction of trajectories when $0<B<1/5$, and the dotted
arrows show them for $-1<B<0$.  First consider the former case,
with $0<B<1/5$.  Expanding cosmologies again have $x>0$, and so again
must be confined to regions $i$ and $iv$.  Now all trajectories in
region $i$ begin at point $2$ and end at point $3$.
Trajectories in region $iv$ are attracted to point $1$ or $3$
in the future, and in the past can be seen to approach $\theta_c$.  If
the stationary point $5$ exists, then trajectories in this region
will approach it in the past.  The collapsing
cosmologies in regions $ii$ and $iii$, can again be seen
to behave in a similar fashion to the expanding solutions, but with
the direction of time reversed.  The cases in which $-1<B<0$, depicted by
dotted arrows, have exactly the opposite behaviour to the description
just given.  Past and future attractors must then be interchanged, but
the description is otherwise the same.

Let us now consider further the asymptotic form of the scale factor as the
stationary points are approached. The functional form of the
scale-factor, $a(t)$, at points $5$ and
$6$ has been determined above, and is given by (\ref{vacuum}).  If
\begin{equation}
\label{range}
\frac{B_5 (1+B_5)}{(1-3 B_5+2 B_5^2)}>0
\end{equation}
then the late-time behaviour of the expanding point $5$
approaches a simple scaling solution as $t \rightarrow \infty$.
If this condition on $B_5$ is not met, then in order to maintain $H>0$
it must be the case that $t<t_0$.  In the limit that $t\rightarrow t_0$ the
scale factor then diverges, and a big-rip singularity occurs.  It can
be seen that divergent behaviour occurs if $-1<B_5<0$, or $1/2<B_5<1$.  This
range of $B$ contains some interesting $f(R)$ theories, as we will
show later on.  Again, the behaviour of $a(t)$ as point $6$ is
approached is similar, but with the direction of time reversed.

The points $3$ and $4$ have a simpler interpretation,
as, independent of $B$, the scale factor evolves like a radiation dominated flat Friedmann
universe when they are approached.  Point $3$ always
corresponds to an expanding universe.  For $0<B<1/5$ this expanding
solution is stable, and trajectories that are attracted toward it
then approach the simple scaling behaviour (\ref{radiation}) as $t
\rightarrow \infty$.  For $B<0$ or $B>1/5$ the expanding solution at
$3$ is a repellor.  Trajectories that originate from this point
therefore have a big-bang singularity in their past, which they approach as
$a \propto (t-t_0)^{\frac{1}{2}}$ when $t \rightarrow t_0$.  The
collapsing trajectories corresponding to point $4$ have a similar
interpretation, with the direction of time reversed.

Finally, consider the points $1$ and $2$, where
$x/y \rightarrow 0$ and $1/y
\rightarrow 0$.  These points correspond to
$a^{\prime}/a \rightarrow$0, which is either a stationary point in the
evolution of $a(t)$, an approach to Minkowski space or a divergence in
$R$.  In order to determine which of these is the case it
appears necessary to have a knowledge of the functional form of $f(R)$, so that equations
(\ref{vac1})-(\ref{vaccon}) can be solved for in these limits.  We
can, however, make some progress in the cases in
which $A_{1,2}$ and $B_{1,2} \rightarrow$constant. Equations (\ref{vac2}),
(\ref{vac3}) and (\ref{vaccon}) can then be integrated to give
\begin{align}
x &\rightarrow \frac{1}{6} Q B (1+A) (\tau-\tau_0 )\\
y &\rightarrow \frac{1}{(1+A) (\tau-\tau_0)}\\
z &\rightarrow z_0 +\frac{2 A}{(1+A)} \ln (\tau -\tau_0 ),
\end{align}
where $\tau_0$ and $z_0$ are constants.  The definition
$d\tau=\sqrt{\vert R \vert} dt$ can also be integrated to give
\begin{equation}
(\tau-\tau_0) \propto (t-t_0)^{(1+A)}.
\end{equation}
It can now be seen that $y \rightarrow \pm \infty$  corresponds to
$\tau \rightarrow \tau_0$.  For $A>-1$ this
corresponds to $t \rightarrow t_0$, while for $A<-1$ it corresponds to
$t\rightarrow \pm \infty$.  The scale factor is now found to be
\begin{equation}
\label{abounce}
\ln \left( \frac{a}{a_0} \right) = (\tau-\tau_0)^2 \propto (t-t_0)^{2 (1+A)},
\end{equation}
where $a_0$ is a constant.  A solution that
approaches $\theta_{1,2}$ with $A>-1$ reaches a stationary point in its
expansion at a finite time, $t_0$.  This solution must
then be matched onto another solution with a similar stationary
point in its past.   If
$A<-1$ as $\theta_{1,2}$ is approached, then these points are reached
only as $t\rightarrow \infty$.  These trajectories then approach Minkowski
space asymptotically.

\section{Perfect Fluid Cosmologies}

\subsection{The Dynamical System}

Now consider the case $\rho \neq 0$ in equations (\ref{Friedmann1})-(\ref{Ricci}).
In this case we begin by transforming the time coordinate from
$t$ to $T$ via
\begin{equation}
\label{T}
dT \equiv \sqrt{\frac{8 \pi \rho}{3 \vert f_R \vert}} dt,
\end{equation}
where it has been assumed $\rho \geqslant 0$, so that $T$ is always real and
increases monotonically in $t$.  Defining the new variables $w \equiv
\sqrt{\vert f_R \vert \vert R \vert/8 \pi \rho}$, $x \equiv a^{\prime}/a$,
$y \equiv f_R^{\prime}/f_R$ and $z \equiv \ln \vert R \vert$
the equations (\ref{Friedmann1})-(\ref{Ricci}) can then be written as
\begin{align}
\label{dyn3}
w^{\prime} &=w \left(\left( \frac{1}{2} +A \right) y+\frac{3}{2} \gamma x \right)\\
x^{\prime} &= \left( \frac{3}{2} \gamma -2 \right) x^2 + \frac{1}{2} x
y+\frac{1}{2} Q w^2
\label{dyn1}\\
y^{\prime} &= \left( \frac{3}{2} \gamma+1 \right) x y-\frac{1}{2} y^2 -Q w^2
- 3 \gamma S+4 x^2
\label{dyn2}\\
z^{\prime} &= 2 A y,
\label{dyn4}
\end{align}
with the constraint equation
\begin{equation}
\label{constraint}
x^2+x y -\frac{1}{2} Q B w^2 -S=0,
\end{equation}
where primes are now understood to denote differentiation
with respect to $T$, and $Q$ and $S$ are the signs of $R$ and $f_R$,
respectively.  The functionals $A=A(z)$, and $B=B(z)$ are the same
as before.

It can again be seen that the surfaces $R=0$ are invariant
sub-manifolds of the phase space.  These surfaces are now given by
(\ref{constraint}) as $x^2+x y-S=0$, and it can be seen that
\begin{equation}
(x^2+xy-S)^{\prime} = (x^2+xy-S) (3 \gamma x +y/B),
\end{equation}
showing that there exist no trajectories in this space along which
$R$ changes sign, justifying our choice of $z$.

To find the stationary points, at finite distances in $x$ and $y$,
we can eliminate $w$ using the constraint equation, 
(\ref{constraint}).  The dynamical system is then described by a
three-dimensional closed system of equations, with four stationary points at finite
distance in the ($x$,$y$) plane.  These points are at
\begin{equation}
\label{xy12}
x_{7,8} = \frac{\pm \sqrt{S}}{\sqrt{5-3 \gamma }} \qquad \qquad \qquad \qquad
y_{7,8} = \pm \frac{(4-3\gamma )}{\sqrt{5-3\gamma}} \sqrt{S}
\end{equation}
and
\begin{equation}
\label{xy34}
x_{9,10} = \frac{\pm \sqrt{2 S}}{\sqrt{2-3 \gamma B_0^2-B_0(4+3 \gamma)}}
\qquad \qquad
y_{9,10} =\frac{\mp 3 \gamma B_0 \sqrt{2 S}}{\sqrt{2 -3 \gamma B_0^2 -B_0
    (4+3 \gamma)}},
\end{equation}
where the $\pm$ signs should be chosen consistently for each point,
and where $B \rightarrow B_0=$constant.  If the limit $B
\rightarrow B_0=$constant does not occur, then points $9$ and $10$ do
not exist.  These points must also have real $x_i$ and $y_i$ in order
to exist.

We will now find the form of the scale factor for the points
$7$-$10$.  From the definitions of $x$ and $y$ it
is clear that at $x=x_i$ and $y=y_i$ we
should have $f_{R}\propto e^{y_i (T-T_0)}$, and $a\propto e^{x_i
  (T-T_0)}$.  The conservation equation (\ref{conservation}) then
yields $\rho \propto a^{-3 \gamma} \propto e^{-3 \gamma x_i
  (T-T_0)}$, which, for $B\neq 1$, allows (\ref{T}) to be integrated to
\begin{equation}
(t-t_0) \propto e^{\frac{1}{2} (y_i+3 \gamma x_i) (T-T_0)}.
\end{equation}
The scale factor can then be written, as a function of $t$, in the
power law form
\begin{equation}
a_i \propto (t-t_0)^{\frac{2}{3 \gamma +y_i/x_i}}.
\end{equation}
At points $7$ and $8$ the scale factor can be seen
to evolve like the Tolman solution, (\ref{radiation}), while at
points $9$ and $10$ it evolves like the
matter dominated solution, (\ref{matter}).  One may expect a different
behaviour at these points when $B=1$, but for this value of $B$ points $9$
and $10$ do not exist for any positive $\gamma$.

Now consider stationary points at infinite distances in the
($x$,$y$) plane.  To identify these points it is convenient to
transform to polar coordinates via $x
\equiv \hat{r} \cos \phi$ and $y \equiv \hat{r} \sin \phi$.
Re-defining the radial coordinate as $\hat{r}\equiv r/(1-r)$
maps the infinite ($x$,$y$) plane to a unit disk, where
$r\rightarrow 1$ in the limit $\hat{r}\rightarrow \infty$.  In these
coordinates equations (\ref{dyn1}) and (\ref{dyn2}) become
\begin{align*}
r^{\prime} &= \frac{1}{4} \Big[ (6 \gamma-5) r^2 \cos \phi -3 \cos (3
  \phi) r^2 -3(4 \gamma S (1-r)^2 -r^2) \sin \phi\\
&\qquad \qquad \qquad \qquad \qquad \qquad  +5r^2 \sin (3 \phi) +2 Q (1-r)^2 (\cos \phi-2 \sin \phi) w^2 \Big]\\
\phi^{\prime} &= \frac{1}{2r(1-r)} \Big[ \cos \phi (3 r^2-6 \gamma S
  (1-r)^2+5 r^2 \cos (2\phi )\\
&\qquad \qquad \qquad \qquad \qquad \qquad  +3 r^2 \sin (2\phi
  )) -Q (1-r)^2 (2 \cos \phi + \sin \phi) w^2 \Big],
\end{align*}
where $w$ is given by (\ref{constraint}) as
\begin{equation}
w^2 = \frac{(r^2-2 S (1-r)^2 +r^2 \cos(2\phi) +r^2
  \sin(2\phi))}{B Q (1-r)^2}.
\end{equation}
As $r\rightarrow 1$, and infinite distances in the ($x$,$y$) plane are
approached, these equations become
\begin{align}
\label{rprime}
r^{\prime} &\rightarrow \frac{1}{4 B} \Big[ (1-(5-6 \gamma)B) \cos
  \phi + 3 (1-B) \cos (3 \phi) - (1-3 B) \sin \phi - (1-5 B)\sin
  (3\phi) \Big]\\
\phi^{\prime} &\rightarrow - \frac{(1+\cos (2\phi )+\sin (2 \phi ))}{2 B (1-r)} \Big[ (1+B) \sin
  \phi+2 (1-2 B) \cos \phi \Big],
\label{phi2} 
\end{align}
independent of the behaviour of $B$.  It can now be seen from
(\ref{phi2}) that stationary points at infinite distances, as $r \rightarrow 1$, can exist
only if $1+\cos (2 \phi )+ \sin (2 \phi ) =0$, or $\tan \phi = 2 (2
B-1)/(B+1)$.  This gives stationary points at 
\begin{equation}
\phi_1=-\pi/2, \qquad \phi_2=\pi/2, \qquad \phi_3=-\pi/4, \qquad
 \phi_4=3\pi/4,
\end{equation}
 and at the two
solutions of 
\begin{equation}
\tan \phi_{5,6} = 2 (2B_0-1)/(B_0+1)
\end{equation}
that exist in the range ($0$,$2\pi$).  These points can be seen to be
in similar positions to those in the vacuum cosmologies, and we will see that
the scale factor evolves in a similar way as they are approached.
Points $1$ to $4$ exist for any $B$, while $5$ and
$6$ require $B\rightarrow B_0=$constant as they are approached, in
order to be stationary.

To find the evolution of the scale factor at points $3$-$6$
we first see that the definitions of $x$ and $y$ allow an integral
such that $f_R \propto a^{\tan \phi_i}$, just as in the vacuum FRW case.
Substituting this into (\ref{dyn1}) and (\ref{dyn2}) allows us to
obtain
\begin{equation}
\frac{a^{\prime\prime}}{a} \rightarrow
\frac{((3\gamma+2)+\frac{3}{2}(2+\gamma)\tan \phi_i-\frac{1}{2} \tan^2
\phi_1)}{(2+\tan \phi_i)} \frac{a^{\prime 2}}{a^2}-\frac{3 \gamma S}{(2+\tan \phi_i)}
\end{equation}
as these points are approached.  This equation can then be integrated to give
\begin{align}
x^2 =\frac{a^{\prime 2}}{a^2} &= \frac{6 \gamma S}{6
  \gamma+(4+3\gamma) \tan \phi_i - \tan^2\phi_i} +c_1 a^{3 \gamma+\tan
  \phi_i \frac{(4-\tan \phi_i)}{(2+\tan \phi_i)}}\\
&\rightarrow c_1 a^{3 \gamma+\tan
  \phi_i \frac{(4-\tan \phi_i)}{(2+\tan \phi_i)}},
\end{align}
where the second line is in the limit $x^2 \rightarrow 
\infty$, as is the case for all of points $3$-$6$.  This
equation can be integrated again, for $\tan \phi_i \neq 0$ or $1$, to find the form of the scale
factor, as points $3$-$6$ are approached, to be
\begin{align}
a &\propto (T-T_0)^{-\frac{2 (2+\tan \phi_i)}{(6 \gamma+(4+3 \gamma)
    \tan \phi_i-\tan^2 \phi_i)}} \\
&\propto (t-t_0)^{\frac{2 \cot \phi_i+1}{\tan \phi_i -1}},
\end{align}
where $T_0$ and $t_0$ are constants of integration, and the second
line has been obtained by an integration of equation (\ref{T}), the
definition of $T$.  Substituting $\phi_i$ into the expression above
shows that points $3$ and $4$ correspond to the Tolman solution,
(\ref{radiation}), and points $5$ and $6$ to the vacuum dominated
solution, (\ref{vacuum}).  This is as may have been expected by analogy to
the vacuum cosmologies, investigated above.  The special cases $\tan
\phi_i =0$ and $1$, corresponding to points $5$ and $6$
when $B_0=1/2$ or $1$, give $a_{5,6} \propto \exp \{c(t-t_0)\}$.
Points $1$ and $2$ correspond to $a^{\prime}/a=0$.

The evolution of the scale factor at the stationary points $1$ to $10$  is
summarised in Table 3.  The symbols in this table refer to those used
in Figures \ref{fluidplot1}-\ref{fluidplot6}.
\begin{table}
\begin{center}
\begin{tabular}{|c|c|c|}
\hline
\textbf{Point} & \textbf{a(t)} & \textbf{Symbol}\\ \hline
1, 2 & $a^{\prime}/a=0$ & \text{Circle}\\ 
3, 4 & Tolman solution, (\ref{radiation}) & \text{Square}\\
5, 6 & Vacuum dominated solution, (\ref{vacuum}) & \text{Triangle}\\
7, 8 & Tolman solution, (\ref{radiation}) & \text{Square}\\ 
9, 10 & Matter dominated solution, (\ref{matter}) & \text{Star}\\ 
\hline 
\end{tabular}
\caption{The evolution of the scale factor, $a(t)$, at the critical
  points $1$-$10$ in perfect fluid filled $f(R)$ cosmologies.  The
  listed symbols correspond to those used in Figures
  \ref{fluidplot1}-\ref{fluidplot6}.}
\end{center}
\end{table}

\subsection{Stability Properties}

Having found all stationary points, at both
finite and infinite distances in the ($x$,$y$) plane, we will now proceed to establish their
stability properties.  This will allow some insight into the degree to
which they can be considered the asymptotic limits of the general
solutions to the Friedmann equations, (\ref{Friedmann1}) and (\ref{Friedmann2}).

First consider the points at finite distances in ($x$,$y$), points
$7$-$10$. As these points are approached $z$
can be seen to diverge to $\pm \infty$.  We can therefore check
stability in the ($x$,$y$) plane by perturbing $x$
and $y$ as 
\begin{equation}
\label{perts}
x \rightarrow x_i +u \qquad \qquad \text{and} \qquad \qquad y \rightarrow y_i +v, 
\end{equation}
and checking the signs of the eigenvalues, $\lambda_i$, of the linearised equations
\begin{equation}
u^{\prime} = \lambda_i u \qquad \qquad \text{and} \qquad \qquad
v^{\prime} = \lambda_i v.
\end{equation}
Substituting (\ref{perts}) into (\ref{dyn1}) and (\ref{dyn2}) gives
the linearised system
\begin{align}
\label{pert1}
u^{\prime} &= \frac{(2 (2x_i+y_i)+((6 \gamma-8) x_i+y_i)B)}{2B} u+ \frac{x_i (2+B)}{2B}v\\
v^{\prime} &= \frac{((16 x_i+2 y_i+3 \gamma y_i) B-4(2x_i+y_i))}{2B}u+
\frac{(((2+3 \gamma) x_i-2 y_i)B-4x_i)}{2B}v,
\label{pert2}
\end{align}
where $B$ in these expressions is in the limit
$R \rightarrow 0$ or $\infty$, whichever is appropriate.  The
eigenvalues, $\lambda_i$, are then given by the roots of a quadratic
equation of the form
$\lambda_i^2 +\alpha \lambda_i + \beta =0.$
If $\alpha>0$ and $\beta>0$ we have a stable point in the ($x$,$y$)
plane.  If $\alpha<0$ and $\beta>0$ we have an unstable point, and if
$\beta<0$ we have a saddle point.  The behaviour of $z$ can be found
by checking the sign of $z^{\prime}$.
For points $7$ and $8$ we find that
\begin{equation}
\label{albe1}
\alpha= \mp \frac{((4-3 \gamma)-(5-6\gamma
  )B)\sqrt{S}}{B\sqrt{5-3 \gamma}} \qquad \qquad
  \text{and} \qquad \qquad \beta =
-\frac{(4-3 \gamma
  (1-B))S}{B},
\end{equation}
where the branch of $\mp$ should be chosen consistently with equation
(\ref{xy12}).  Similarly, for points $9$ and $10$ we find
\begin{equation}
\label{albe2}
\alpha = \mp \frac{3 ((1+B) \gamma-2) \sqrt{S}}{\sqrt{2(2-B(4+3
  \gamma (1+B)))}}
\qquad\ \qquad \text{and} \qquad \qquad
\beta=\frac{(4-3 \gamma
  (1-B))S}{B},
\end{equation}
with $\mp$ chosen consistent with (\ref{xy34}).  It can
immediately be seen that one of these two sets of points always
corresponds to a pair of saddles, as $\beta$ has opposite signs
in (\ref{albe1}) and (\ref{albe2}).  The other set can then be seen to contain one
attractor and one repellor, as $\alpha$ has a different sign for each
point in both (\ref{albe1}) and (\ref{albe2}).  Which two points are saddles, which is the
attractor and which is the repellor depends on the values of $B$,
$\gamma$ and $S$.  We will consider various cases below.

Now consider the stability of the stationary points at infinite
distances in ($x$,$y$).  In this case we perturb $\phi$ so
that $\phi \rightarrow \phi_i +\delta \phi$.  For points $1$ and
$2$ we must also consider a perturbation to $z$ such that $z
\rightarrow z_i + \delta z$, and $B \rightarrow B + (dB/dz) \delta z$.
The evolution equation (\ref{rprime}) and (\ref{phi2}) then become
\begin{equation}
\delta \phi^{\prime}_{1,2} = \mp \frac{(1+B)}{(1-r) B} \delta \phi_{1,2}
\qquad \qquad \text{and} \qquad \qquad
r^{\prime}_{1,2} = \pm \frac{1}{2},
\end{equation}
independent of $\delta z$.  The upper sign here is for point $1$, and
the lower sign for point $2$.  These points are considered stable if $r^{\prime}>0$
and $\delta \phi/\phi <0$, unstable if $r^{\prime}<0$ and $\delta
\phi/\phi >0$, and as saddle points otherwise.  Therefore, if $B>0$ or
$<-1$ then one of these points is stable and the other unstable.  If
$-1<B<0$ then both points are saddles.

For points $3$-$6$ both $r$ and $z \rightarrow \infty$, while
$\phi$ is finite (for $B\neq 1/2$ or $1$).  It is therefore sufficient to perturb $\phi$ as
$\phi \rightarrow \phi_i +\delta \phi$, and to check the signs of $r$
and $z$.  As $\phi \rightarrow \phi_i$, and $r \rightarrow \infty$, we
then have for points $\phi_{3,4}$
\begin{equation}
\delta \phi^{\prime}_{3,4} = \mp \frac{(1-5B)}{\sqrt{2} B} \frac{\delta \phi_{3,4}}{(1-r)}
\qquad \qquad \text{and} \qquad \qquad
r^{\prime}_{3,4} = \mp \frac{(5-3\gamma )}{2 \sqrt{2}},
\end{equation}
where the upper sign is for $\phi_3$, and the lower sign for
$\phi_4$.  For points $5$ and $6$
\begin{equation}
\delta \phi^{\prime}_{5,6} = \pm \frac{(1-5 B)}{B \sqrt{1+\frac{4
      (1-2B)^2}{(1+B)^2}}} \frac{\delta \phi_{5,6}}{(1-r)}
\qquad \qquad \text{and} \qquad \qquad
r^{\prime}_{5,6} = \mp \frac{(2-(4+3 \gamma (1+B))B)}{2B(1+B)\sqrt{1+\frac{4 (1-2B)^2}{(1+B)^2}}},
\end{equation}
where the upper sign is for point $5$, and the lower for point $6$.  It
can be seen from the $\delta \phi$ equations that for $0<B<1/5$ the only points that can be stable
are $3$ and $6$, and that the only points that can be
unstable are $4$ and $5$.  For $B<0$ or $B>1/5$ this
behaviour is reversed and $4$ and $5$ are the only points
that can be stable, while $3$ and $6$ are the only
unstable points.  The special cases $B=1/2$ and $1$ can have $z$
asymptoting to a finite value, and so in this case the stability analysis must
include a perturbation to $z$.  In Appendix
\ref{appendix} we perform this analysis, and find similar behaviour to
the corresponding vacuum cosmologies investigated above.

The stability of the critical points $1$-$10$ can be seen to be
strongly dependent on the value of the parameter $B$, as in the vacuum
case.  Now, however, the additional parameters $\gamma$ and
$S$ are also involved in determining stability.  The results for the
different cases $S= \pm 1$, and $\gamma \gtrless 5/3$ are summarised in Tables
\ref{table4}-\ref{table7}.  Having determined the stability properties of all stationary points, at both
finite and infinite distances in the ($x$,$y$) plane, we will now
proceed to investigate what this can tell us about the behaviour of
the general solution.

\begin{table}
\begin{center}
\begin{tabular}{|c|c|c|c|c|}
\hline
\multirow{2}{*}{\textbf{Point}} &
\multicolumn{4}{|c|}{\bf{Stability when $\mathbf{S=-1}$ and $\mathbf{\gamma >5/3}$}} \\[3pt] \cline{2-5}
& $\mathbf{B<-1}$ & $\mathbf{-1<B<0}$ & $\mathbf{0<B<1/5}$ & $\mathbf{1/5<B}$ \\ \hline
1 & A & S & A & A \\[2pt]
2 & R & S & R & R  \\[2pt]
3 & S & S & A & S \\[2pt]
4 & S & S & R & S  \\
\multirow{2}{*}{5} & \scriptsize{$m<0$: A} & \multirow{2}{*}{A} & \scriptsize{$m<0$: S} & \multirow{2}{*}{A}\\[-5pt]
& \scriptsize{$m>0$: S} & & \scriptsize{$m>0$: R} & \\
\multirow{2}{*}{6} & \scriptsize{$m<0$: R} & \multirow{2}{*}{R} & \scriptsize{$m<0$: S} & \multirow{2}{*}{R}\\[-5pt]
& \scriptsize{$m>0$: S} & & \scriptsize{$m>0$: A} & \\
\multirow{2}{*}{7} & \multirow{2}{*}{R} & \multirow{2}{*}{R} &
\multirow{2}{*}{S} & \scriptsize{$n<0$: S}\\[-5pt]
& & & &  \scriptsize{$n>0$: R}\\
\multirow{2}{*}{8} & \multirow{2}{*}{A} & \multirow{2}{*}{A} &
\multirow{2}{*}{S} & \scriptsize{$n<0$: S}\\[-5pt]
& & & &  \scriptsize{$n>0$: A}\\
\multirow{2}{*}{9} & \scriptsize{$m<0$: S} & \multirow{2}{*}{--} &
\scriptsize{$m<0$: R} & \scriptsize{$n<0$: R}\\[-5pt]
&\scriptsize{$m>0$: --} & & \scriptsize{$m>0$: --} &
\scriptsize{$n>0$: S}\\
\multirow{2}{*}{10} & \scriptsize{$m<0$: S} & \multirow{2}{*}{--} &
\scriptsize{$m<0$: A} & \scriptsize{$n<0$: A}\\[-5pt]
&\scriptsize{$m>0$: --} & & \scriptsize{$m>0$: --} &  \scriptsize{$n>0$: S}\\[2pt]
 \hline
\end{tabular}
\end{center}
\caption{The stability properties of the critical points
  $1$-$10$ in perfect fluid $f(R)$ cosmologies, when $S=-1$ and
  $\gamma >5/3$. Here `A' denotes a stable
  Attractor, `R' an unstable Repellor and `S' a Saddle point.  The
  quantities $m$ and $n$ are defined as $m\equiv 2-3 \gamma B^2-B (4+3
  \gamma)$, and $n \equiv 3 \gamma B^2+B(4-3 \gamma )$.  Dashes
  indicate that a point is not present for that range of
  $B$.}
\label{table4}
\end{table}

\begin{table}
\begin{center}
\begin{tabular}{|c|c|c|c|c|}
\hline
\multirow{2}{*}{\textbf{Point}} &
\multicolumn{4}{|c|}{\bf{Stability when $\mathbf{S=-1}$ and
    $\mathbf{0< \gamma <5/3}$}} \\[3pt] \cline{2-5}
& $\mathbf{B<-1}$ & $\mathbf{-1<B<0}$ & $\mathbf{0<B<1/5}$ &
$\mathbf{1/5<B}$ \\ \hline
1 & A & S & A & A \\[2pt]
2 & R & S & R & R  \\[2pt]
3 & R & R & S & R \\[2pt]
4 & A & A & S & A  \\
\multirow{2}{*}{5} & \scriptsize{$m<0$: A} & \multirow{2}{*}{A} & \multirow{2}{*}{R} & \scriptsize{$m<0$: A}\\[-5pt]
& \scriptsize{$m>0$: S} & & & \scriptsize{$m>0$: S}  \\
\multirow{2}{*}{6} & \scriptsize{$m<0$: R} & \multirow{2}{*}{R} & \multirow{2}{*}{A} & \scriptsize{$m<0$: R}\\[-5pt]
& \scriptsize{$m>0$: S} & & & \scriptsize{$m>0$: S}  \\[2pt]
7 & -- & -- & -- & -- \\
8 & -- & -- & -- & -- \\[-5pt]
\multirow{2}{*}{9} & \scriptsize{$m<0$: S} & \multirow{2}{*}{--} &
\multirow{2}{*}{--} & \scriptsize{$m<0$: S}\\[-5pt]
&\scriptsize{$m>0$: --} & &  &
\scriptsize{$m>0$: --}\\
\multirow{2}{*}{10} & \scriptsize{$m<0$: S} & \multirow{2}{*}{--} &
\multirow{2}{*}{--} & \scriptsize{$m<0$: S}\\[-5pt]
&\scriptsize{$m>0$: --} & &  &
\scriptsize{$m>0$: --}\\[2pt]
\hline
\end{tabular}
\end{center}
\caption{The stability properties of the critical points
  $1$-$10$ in perfect fluid $f(R)$ cosmologies, when $S=-1$ and
  $0< \gamma <5/3$. Here A, R, S, and $m$ are defined as in Table \ref{table4}.  Dashes
  again indicate the absence of a point.}
\label{table5}
\end{table}

\begin{table}
\begin{center}
\begin{tabular}{|c|c|c|c|c|}
\hline
\multirow{2}{*}{\textbf{Point}} &
\multicolumn{4}{|c|}{\bf{Stability when $\mathbf{S=1}$ and $\mathbf{\gamma >5/3}$}} \\[3pt] \cline{2-5}
& $\mathbf{B<-1}$ & $\mathbf{-1<B^{\ast}<0}$ & $\mathbf{0<B<1/5}$ &
$\mathbf{1/5<B}$ \\ \hline
1 & A & S & A & A \\[2pt]
2 & R & S & R & R  \\[2pt]
3 & S & S & A & S \\[2pt]
4 & S & S & R & S  \\
\multirow{2}{*}{5} & \scriptsize{$m<0$: A} & \multirow{2}{*}{A} & \scriptsize{$m<0$: S} & \multirow{2}{*}{A} \\[-5pt]
& \scriptsize{$m>0$: S} & & \scriptsize{$m>0$: R} & \\
\multirow{2}{*}{6} & \scriptsize{$m<0$: R} & \multirow{2}{*}{R} & \scriptsize{$m<0$: S} & \multirow{2}{*}{R} \\[-5pt]
& \scriptsize{$m>0$: S} & & \scriptsize{$m>0$: A} & \\
7 & -- & -- & -- & -- \\
8 & -- & -- & -- & -- \\[-5pt]
\multirow{2}{*}{9} & \scriptsize{$m<0$: --} & \multirow{2}{*}{$\; \text{A}^{\dagger}$} &
\scriptsize{$m<0$: --} & \multirow{2}{*}{--} \\[-5pt]
& \scriptsize{$m>0$: A} & & \scriptsize{$m>0$: S} &   \\
\multirow{2}{*}{10} & \scriptsize{$m<0$: --} & \multirow{2}{*}{$\; \text{R}^{\dagger}$} &
\scriptsize{$m<0$: --} & \multirow{2}{*}{--} \\[-5pt]
& \scriptsize{$m>0$: R} & & \scriptsize{$m>0$: S} & \\[2pt]
\hline
\end{tabular}
\end{center}
\caption{The stability properties of the critical points
  $1$-$10$ in perfect fluid $f(R)$ cosmologies, when $S=1$ and
  $\gamma >5/3$. A, R, S, and $m$ are defined as in Table \ref{table4}.  Dashes
  indicate the absence of a point.  The $\dagger$ here indicates
  that these points have the indicated stability as long as $\gamma
  <2$.  The $\ast$ indicates the range of $B$ for which, in this case, there exist
  regions of the phase space without either an attractor or a repellor.}
\label{table6}
\end{table}

\begin{table}
\begin{center}
\begin{tabular}{|c|c|c|c|c|}
\hline
\multirow{2}{*}{\textbf{Point}} &
\multicolumn{4}{|c|}{\bf{Stability when $\mathbf{S=1}$ and $\mathbf{0<\gamma <5/3}$}} \\[3pt] \cline{2-5}
& $\mathbf{B<-1}$ & $\mathbf{-1<B^{\ast}<0}$ & $\mathbf{0<B<1/5}$ &
$\mathbf{1/5<B}$ \\ \hline
1 & A & S & A & A \\[2pt]
2 & R & S & R & R  \\[2pt]
3 & R & R & S & R \\[2pt]
4 & A & A & S & A  \\
\multirow{2}{*}{5} & \scriptsize{$m<0$: A} & \multirow{2}{*}{A} & \multirow{2}{*}{R} & \scriptsize{$m<0$: A} \\[-5pt]
& \scriptsize{$m>0$: S} & & & \scriptsize{$m>0$: S} \\
\multirow{2}{*}{6} & \scriptsize{$m<0$: R} & \multirow{2}{*}{R} & \multirow{2}{*}{A} & \scriptsize{$m<0$: R} \\[-5pt]
& \scriptsize{$m>0$: S} & & & \scriptsize{$m>0$: S} \\
\multirow{3}{*}{7} & \scriptsize{$m<0$: S} & & & \multirow{3}{*}{S} \\[-5pt]
& \scriptsize{$m>0,n<0$: A} & \scriptsize{$n<0$: A} &
\scriptsize{$n<0$: A} & \\[-5pt]
& \scriptsize{$m>0,n>0$: S} & \scriptsize{$n>0$: S} &
\scriptsize{$n>0$: S} &\\
\multirow{3}{*}{8} & \scriptsize{$m<0$: S} & & & \multirow{3}{*}{S} \\[-5pt]
& \scriptsize{$m>0,n<0$: R} & \scriptsize{$n<0$: R} &
\scriptsize{$n<0$: R} & \\[-5pt]
& \scriptsize{$m>0,n>0$: S} & \scriptsize{$n>0$: S} &
\scriptsize{$n>0$: S} &\\
\multirow{2}{*}{9} & \scriptsize{$n<0$: S} & \scriptsize{$n<0$: S} & \scriptsize{$n<0$: S}
& \scriptsize{$m<0$: --} \\[-5pt]
& \scriptsize{$n>0$: A} & \scriptsize{$n>0$: A} & \scriptsize{$n>0$:
  A}& \scriptsize{$m>0$: A} \\
\multirow{2}{*}{10} & \scriptsize{$n<0$: S} & \scriptsize{$n<0$: S} & \scriptsize{$n<0$: S}
& \scriptsize{$m<0$: --} \\[-5pt]
& \scriptsize{$n>0$: R} & \scriptsize{$n>0$: R} & \scriptsize{$n>0$:
  R}& \scriptsize{$m>0$: R} \\[2pt]
\hline
\end{tabular}
\end{center}
\caption{The stability properties of the critical points
  $1$-$10$ in perfect fluid $f(R)$ cosmologies, when $S=1$ and
  $0< \gamma <5/3$. A, R, S, and $m$ are defined as in Table \ref{table4}.  Dashes
  indicate the absence of a point.}
\label{table7}
\end{table}

\subsection{The General Solution}

The general solutions to the Friedmann equations
(\ref{Friedmann1})-(\ref{Friedmann2}) in the presence of a perfect
fluid are more complicated than the vacuum case.  The
phase space of solutions is now, in general, three dimensional.  Nevertheless, it is
still possible to make progress in understanding the general
solution.

\begin{figure}[htb]
\begin{center}
\subfigure[$S=-1$]{\epsfig{figure=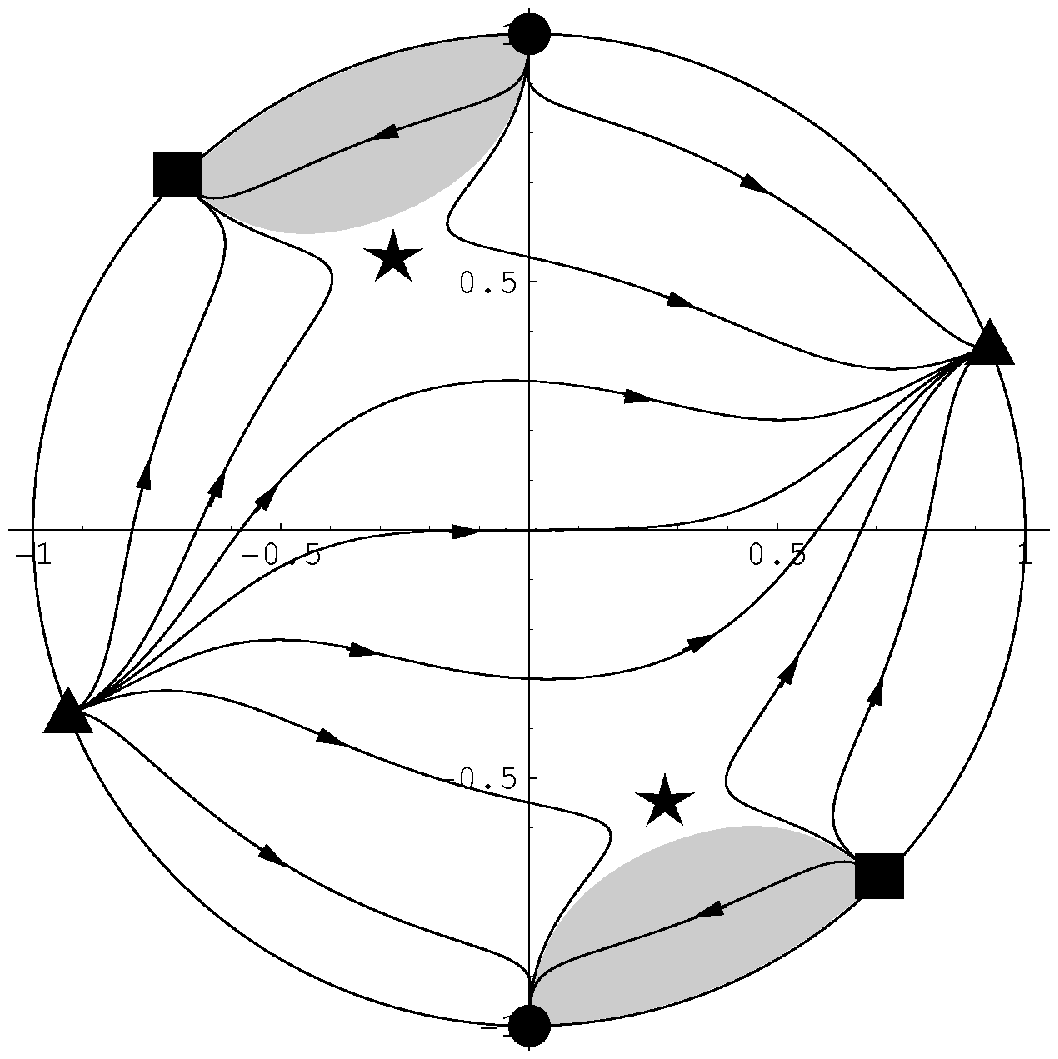,height=7cm}}
\qquad
\subfigure[$S=1$]{\epsfig{figure=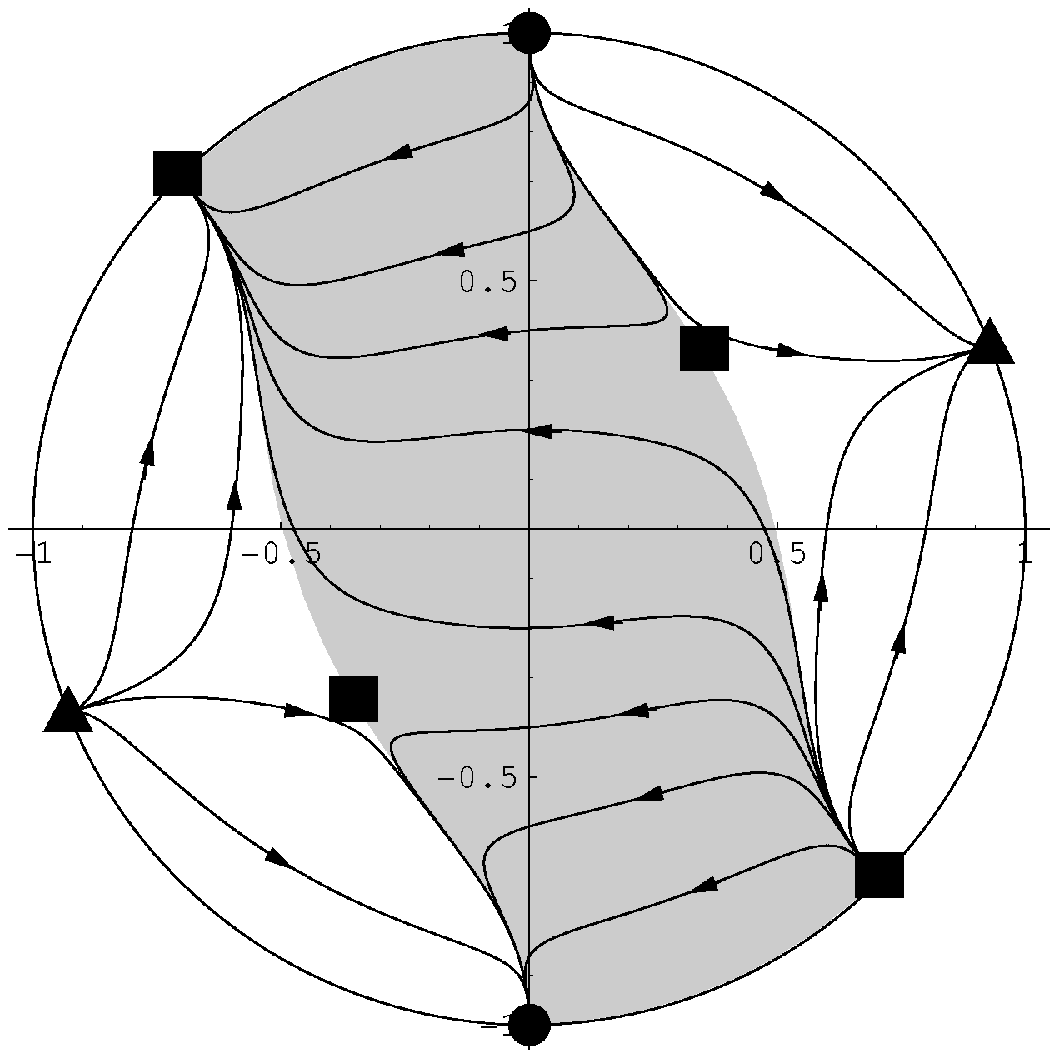,height=7cm}}
\caption{The ($x$,$y$) plane in the phase space of solutions defined
  by equations (\ref{dyn3})-(\ref{constraint}).  Here the infinite ($x$,$y$) plane has been
  compacted to a unit disk.  Dots, squares, triangles and stars indicate the position of stationary
  points.  Squares correspond to points $3$, $4$, $7$ and $8$, and
  hence the Tolman solution, (\ref{radiation}).  Stars
  correspond to points $9$ and $10$, and the matter dominated solution,
  (\ref{matter}).  Triangles are at points $5$ and
  $6$, corresponding to the vacuum dominated solution,
  (\ref{vacuum}).  Dots are at points $1$ and $2$, where
  $a^{\prime}/a=0$.  The phase plane is separated into three different
  regions by the invariant sub-manifolds where $R=0$.  Grey
  regions have $R<0$, and white regions have $R>0$.  In these plots we
  take $B=2/3$ and $\gamma=1$, corresponding to pressureless dust and a
  gravitational Lagrangian dominated by a term $\sim R^3$.}
\label{fluidplot1}
\end{center}
\end{figure}

\begin{figure}[htb]
\begin{center}
\subfigure[$S=-1$]{\epsfig{figure=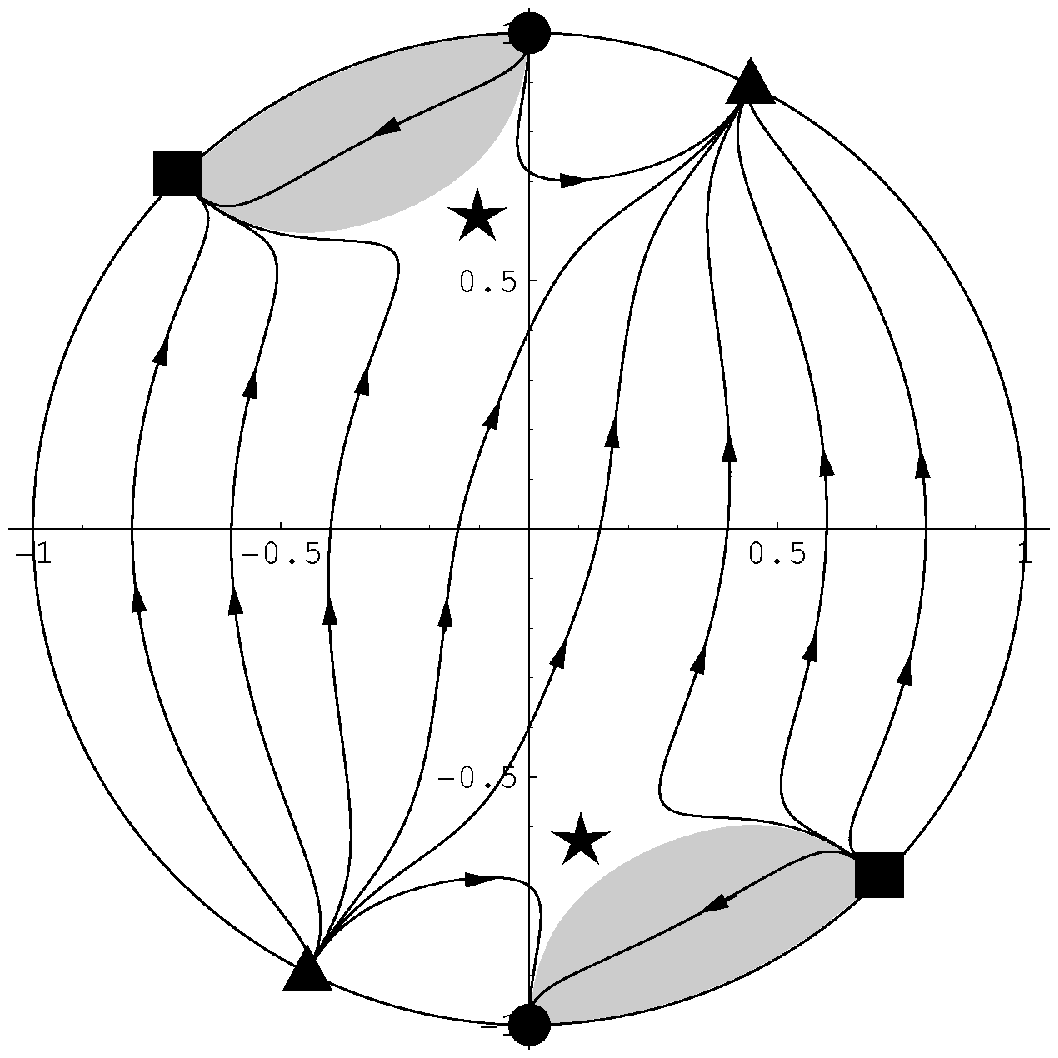,height=7cm}}
\qquad
\subfigure[$S=1$]{\epsfig{figure=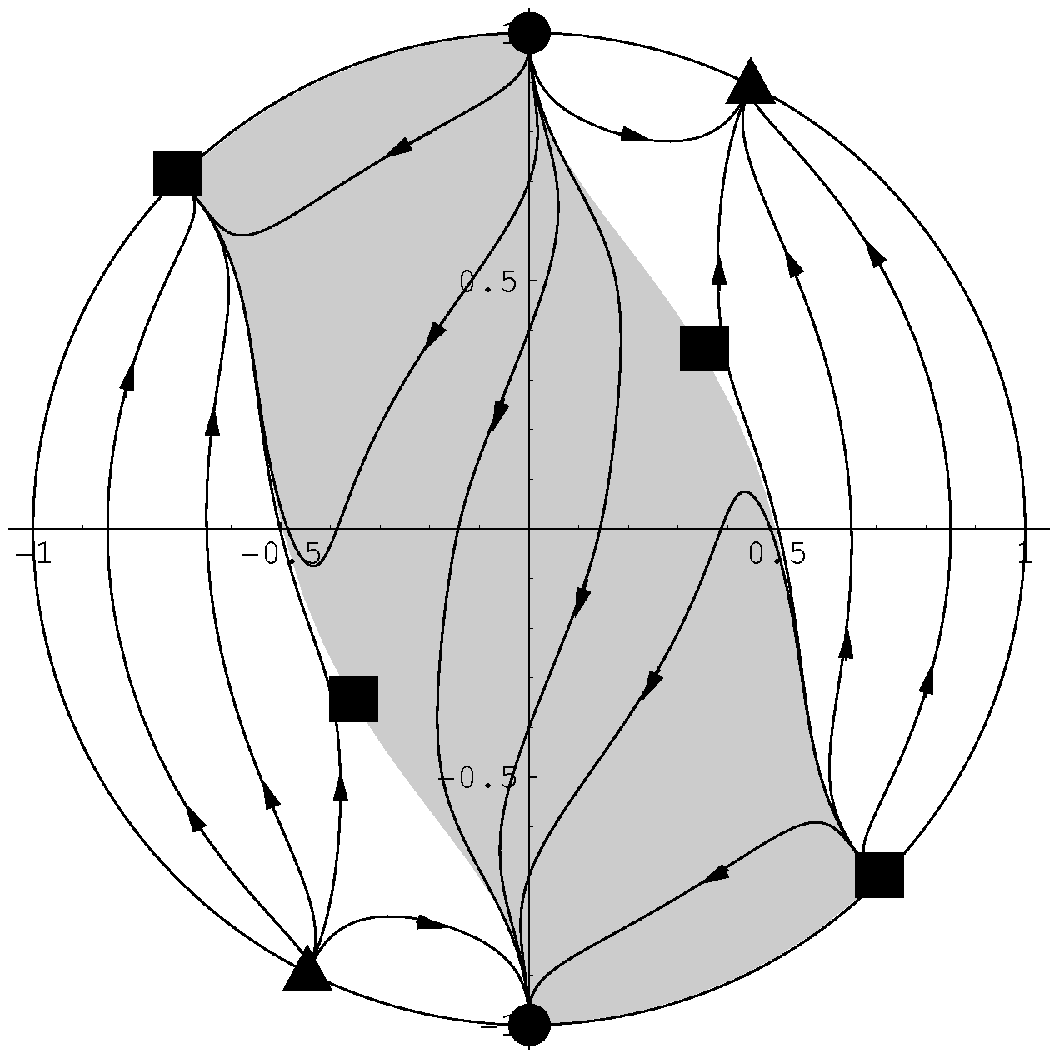,height=7cm}}
\caption{These plots show the ($x$,$y$) plane in the phase space of
  solutions when $B=2$ and $\gamma=1$, corresponding to pressureless dust and a
  gravitational Lagrangian dominated by a term $\sim R^{-1}$.  See the
  caption of Figure \ref{fluidplot1} for further details.}
\label{fluidplot2}
\end{center}
\end{figure}

\begin{figure}[htb]
\begin{center}
\subfigure[$S=-1$]{\epsfig{figure=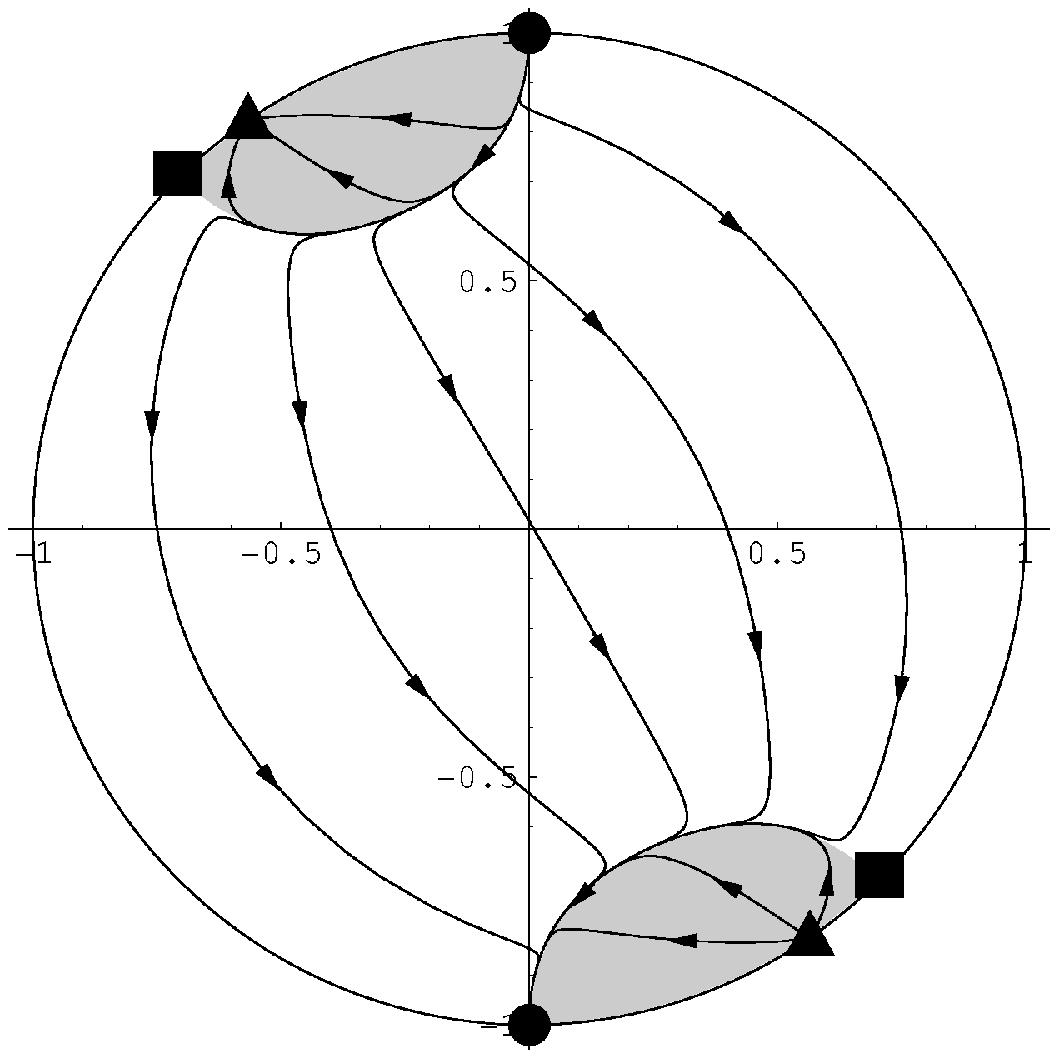,height=7cm}}
\qquad
\subfigure[$S=1$]{\epsfig{figure=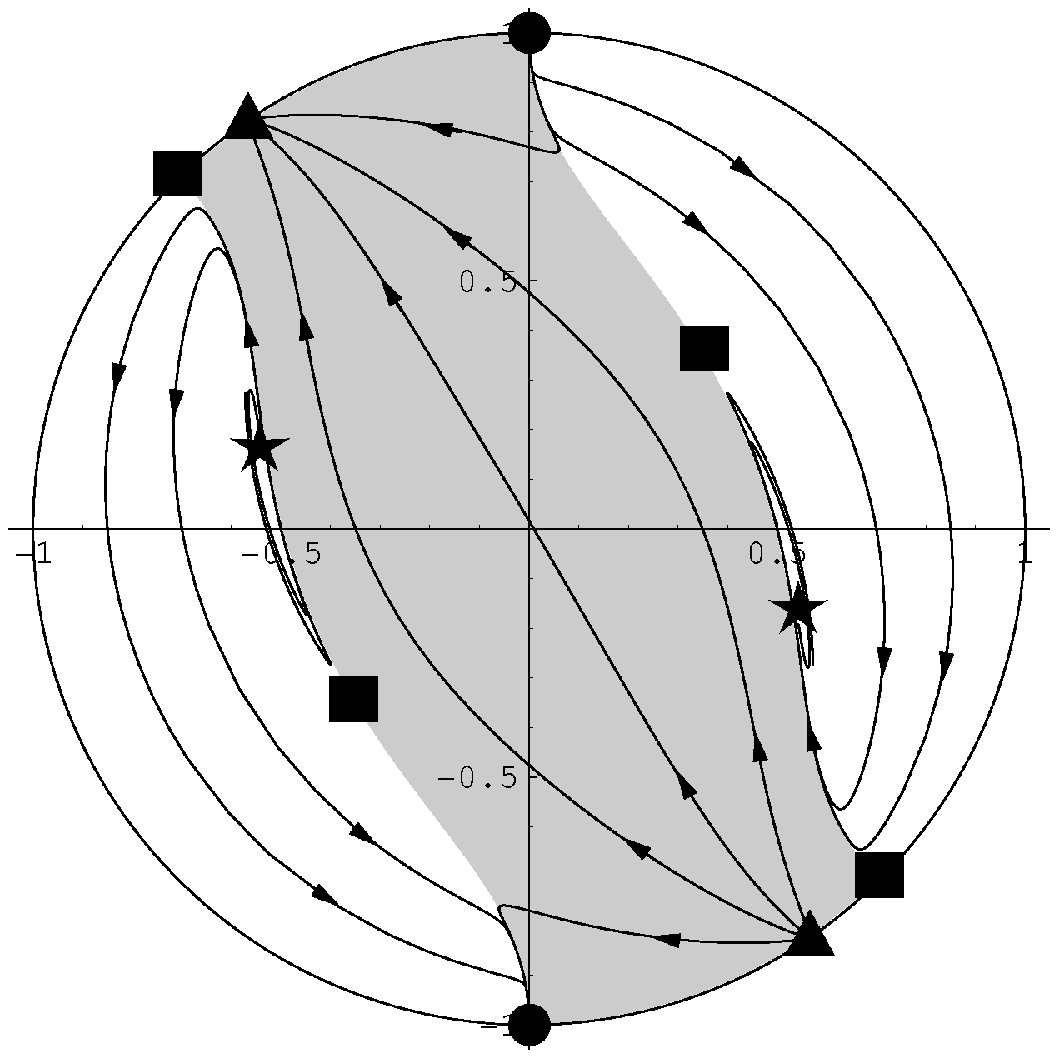,height=7cm}}
\caption{Plots showing the ($x$,$y$) plane in the phase space of
  solutions when $B=1/10$ and $\gamma=1$, corresponding to pressureless dust and a
  gravitational Lagrangian dominated by a term $\sim R^{\frac{10}{9}}$.  See the
  caption of Figure \ref{fluidplot1} for further details.}
\label{fluidplot3}
\end{center}
\end{figure}

\begin{figure}[htb]
\begin{center}
\subfigure[$S=-1$]{\epsfig{figure=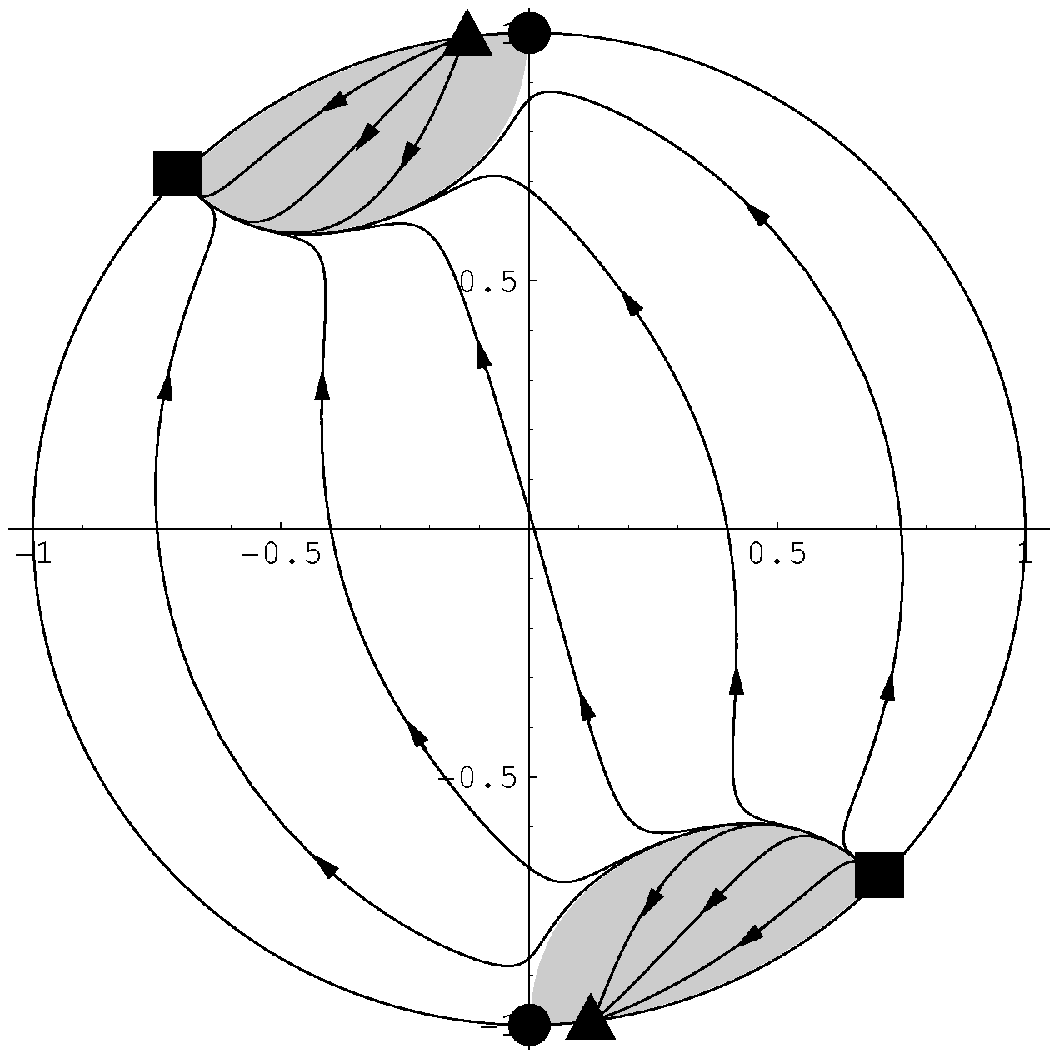,height=7cm}}
\qquad
\subfigure[$S=1$]{\epsfig{figure=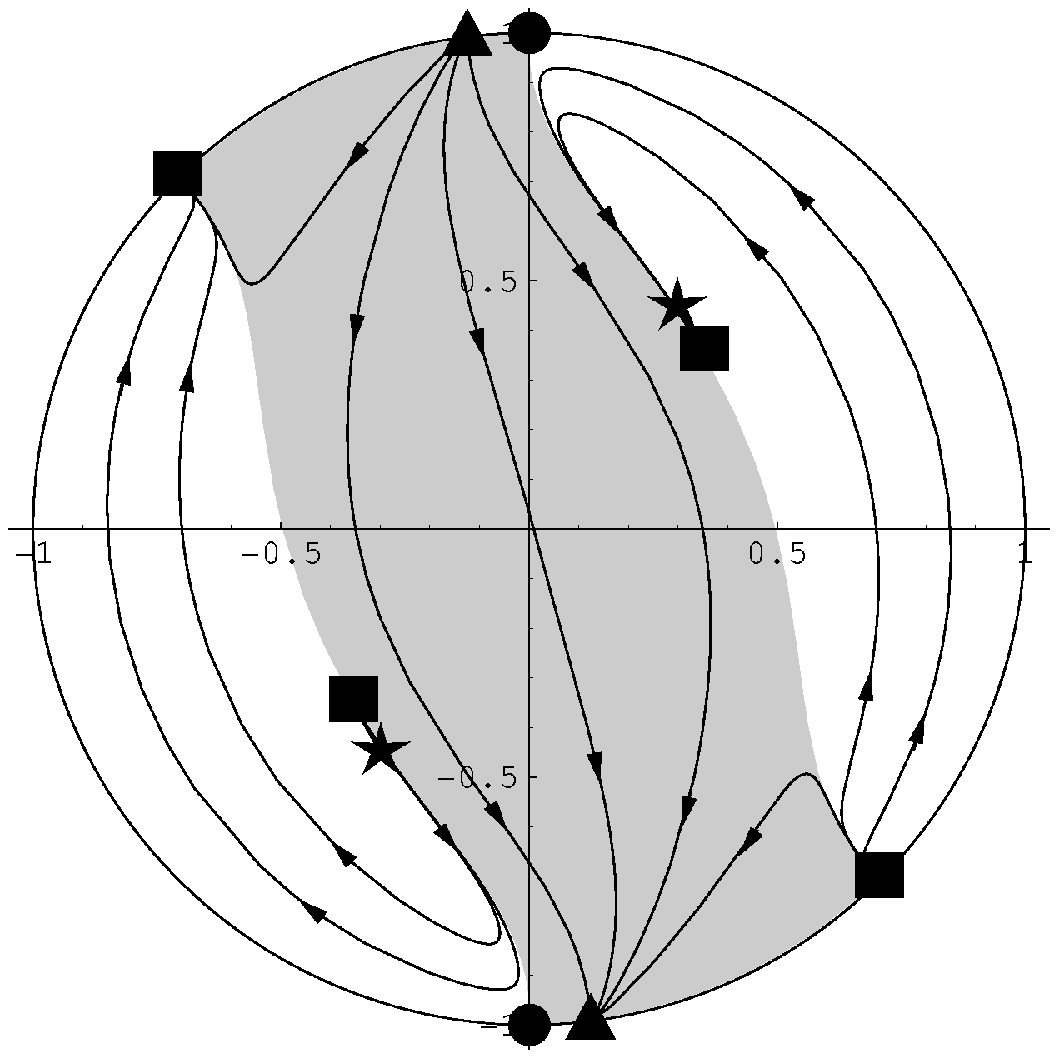,height=7cm}}
\caption{Plots showing the ($x$,$y$) plane in the phase space of
  solutions when $B=-1/2$ and $\gamma=1$, corresponding to pressureless dust and a
  gravitational Lagrangian dominated by a term $\sim R^{\frac{2}{3}}$.  See the
  caption of Figure \ref{fluidplot1} for further details.}
\label{fluidplot5}
\end{center}
\end{figure}

\begin{figure}[htb]
\begin{center}
\epsfig{figure=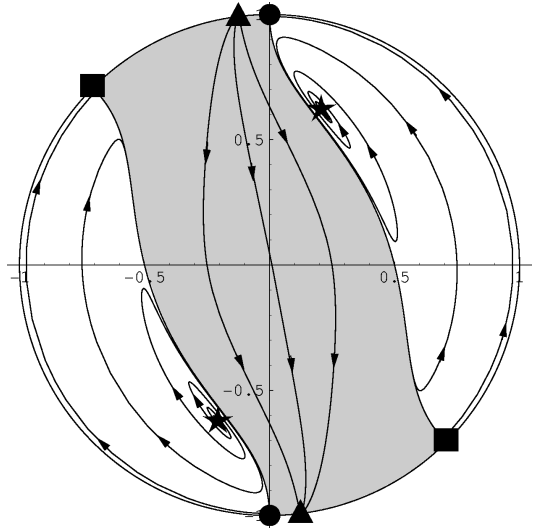,height=7cm}
\caption{Plots showing the ($x$,$y$) plane in the phase space of
  solutions when $B=-1/2$, $S=1$ and $\gamma=2$.  This situation
  corresponds to a stiff fluid and a
  gravitational Lagrangian dominated by a term $\sim
  R^{\frac{2}{3}}$.  The left-hand white region in this plot has only
  a single repellor, and no attractor.  Trajectories in this region
  can then be seen to originate from the repellor, but have no stable
  critical point to attract them in the future.  They therefore spiral
  outwards forever, away from the repellor.  The right-hand white region can
  similarly be seen to have an attractor but no repellor.  See the
  caption of Figure \ref{fluidplot1} for further details.}
\label{fluidplot6}
\end{center}
\end{figure}

As was previously noted, the surfaces $x^2+xy=S$ (corresponding to
$R=0$) are invariant sub-manifolds of the phase space.
This equation has two roots, describing two non-intersecting surfaces
that separate the the phase space into three different regions, between
which trajectories cannot cross.  The shape of these regions
depends on the sign of $S$, and is illustrated in Figures
\ref{fluidplot1}-\ref{fluidplot6} by the grey and white areas.  These plots show the ($x$,$y$)
plane, and some representative trajectories within it, for various
values of $\gamma$ and $B$.  The left-hand plot, (a), in each of
these figures corresponds to $S=-1$, and the right-hand plot, (b), to
$S=1$.

A number of parallels can be drawn between the perfect fluid
cosmologies, illustrated in Figures \ref{fluidplot1}-\ref{fluidplot6},
and the vacuum cosmologies, illustrated in Figure \ref{vacplot}.  The circle
at $r=1$ in Figures \ref{fluidplot1}-\ref{fluidplot6} can be seen to be
quite similar to the circle in Figure \ref{vacplot}.  They both have
stationary points at $\pi/2$ and $-\pi/2$, corresponding to
$a^{\prime}/a=$constant, and at $-\pi/4$ and $3 \pi/4$, corresponding to
$a\sim t^{\frac{1}{2}}$ (although the coordinates have different
definitions in each case).  Whatsmore, in both cases these points mark the
boundaries of regions between which trajectories cannot pass.  A
further similarity is in the position of the triangles, which
correspond to the vacuum dominated solution, (\ref{vacuum}).
In both cases, for $B>1/5$ or $<-1$, these points exist in what are
labelled as region $i$ and $iii$ in Figure \ref{vacplot}, and for $-1<B<1/5$, they
exist in regions $ii$ and $iv$.

All of the points at $r=1$ therefore have analogous points in the
vacuum cosmology case, and all of these points can be seen to exist
for any $\gamma$, $B$ and $S$ (as long as $B \rightarrow B_0=$constant in the case
of points $5$ and $6$).  As $r$ increases these perfect fluid
cosmologies therefore approach the behaviour of the vacuum cosmologies
studied above.  The principle difference between the perfect fluid
case and the vacuum case is
due to the existence of the extra dimension in the phase space.  Points that were
previously attractors (or repellors) in vacuo, can now be
saddles in the presence of a perfect fluid, as they can be
unstable (or stable) in the extra dimension.  Whether these points
maintain the attractor/repellor nature they exhibited in vacuo, or become saddles, can be
read off from Tables \ref{table4}-\ref{table7}.

As well as the points at $r=1$, there can exist four further points at finite
distances in the ($x$,$y$) plane.  These are points $7$-$10$, given by equations
(\ref{xy12}) and (\ref{xy34}).  They have no analogy in the
vacuum cosmologies considered above, and do not exist for all
$\gamma$, $B$ and $S$.  Points $7$ and $8$ correspond to the
Tolman solution, (\ref{radiation}), and exist on the boundaries between grey and
white regions.  Points $9$ and $10$, correspond to the matter
dominated solution (\ref{matter}), and can exist in either the grey or
white regions.  For $S=-1$ they are saddles if they
exist in the the white region, and attractor/repellors if they exist
in the grey regions.  For $S=1$ this behaviour is reversed.  If both
of these sets of points exist then one of them will be a pair of
saddle points, while the other will be an attractor/repellor pair.
In general, the existence and stability of these points are functions
of $\gamma$, $B$ and $S$, the details of which can be read off from
Tables \ref{table4}-\ref{table7}.

The existence, location and stability of the critical points, in the
presence of a perfect fluid, is more complicated than
the vacuum case.  Rather than discuss all possible trajectories, we
will therefore show a number of examples that we consider to be illustrative.
In these examples $B$ will taken as a constant.  It should be
emphasised that in general $B$ is {\it not} a constant.  However, for the
purposes of illustrating the existence and attractor/repellor nature
of the stationary points, for various different values of $B$, it is
convenient to take it as so.

Figure \ref{fluidplot1} shows the ($x$,$y$) plane for a universe filled
with pressureless dust, $\gamma=1$, for the case $B=2/3$.  Such a value
of $B$ may be expected if $f(R) \sim R^3$, and so may be of interest
when higher powers of the curvature are expected to dominate the
gravitational Lagrangian.  Here the points at $r=1$ have locations and
stability analogous to the vacuum case.  There are also two points at
finite distance, corresponding to the matter dominated solution,
(\ref{matter}), when $S=-1$, and the Tolman solution,
(\ref{radiation}), when $S=1$.  These points, however, are both saddles,
and asymptotic behaviour in this case is therefore very similar to the
corresponding vacuum cosmology.

Consider now the effect of lower orders of $R$ dominating the
gravitational Lagrangian.  If $f(R) \sim 1/R$ dominates then we may expect $B
\rightarrow 2$.  This case, with a dust equation of state, is
illustrated in Figure \ref{fluidplot2}.  The asymptotes of the
trajectories in Figure \ref{fluidplot2} can be seen to be very similar
to those in Figure \ref{fluidplot1}, and again the stability and location of the critical
points are analogous to the corresponding vacuum cosmology.  Only saddle points are
introduced at finite distances in ($x$,$y$).

Let us now consider some ranges of $B$ for which we know
the stability properties of the critical points to change.  First
consider $B$ in the range $0<B<1/5$, say $B=1/10$.  This case
would be expected from $f(R) \sim R^{\frac{10}{9}}$, and is shown
in Figure \ref{fluidplot3}, for $\gamma =1$.
For $S=-1$ the points 3 and 4, at $-\pi/4$ and $3\pi/4$, are
now no longer attractor/repellors, but saddle points.
Trajectories that would otherwise have ended on them are now therefore
drawn to points 1 and 2, at $-\pi/2$ and $\pi/2$.  There are no new
points at finite distances in this case, and the behaviour is
otherwise the same as the corresponding vacuum cosmology.  For $S=1$
points 3 and 4 are again saddles.  Now, however, all four possible
points at finite distances exist.  The points 9 and 10 are now
attractor/repellors, and so this case is
qualitatively different from the corresponding vacuum cosmology, with
new asymptotes possible.  The case with $S=1$ is that which was
considered in \cite{Power}.

Another range of $B$ which we may wish to consider is $-1<B<0$, which
is shown in
Figure \ref{fluidplot5} for $B=-1/2$, corresponding to $f(R)
\sim R^{\frac{2}{3}}$.  The stability of the triangular points, corresponding to the
vacuum solution (\ref{vacuum}), is now reversed, as would be expected
from analogy with the vacuum cosmology.  Now the circular points at $-\pi/2$
and $\pi/2$ are saddles, and the squares again have the
attractor/repellor behaviour that would be expected from the
corresponding vacuum cosmologies.  The case $S=-1$ contains no
extra critical points at finite distances, and so again has asymptotics that
are otherwise the same as the vacuum case.  For
$S=1$ all four points at finite distances are present, and again
points 9 and 10 act as attractor/repellors.

These examples have been intended to show some representative
illustrations of the ($x$,$y$) phase plane, for different values of
$B$.  The figures shown above do not show all possible behaviours, and
should not be considered exhaustive.  It is, for example, quite
possible to have points 7 and 8 acting as attractor/repellors,
although this was not explicitly the case in any of the examples
above.  The reader will be able to find ranges of $B$ and $\gamma$
where this behaviour occurs, for either $S$, by referencing Tables
\ref{table4}-\ref{table7}, above.

It is of interest to note that for all but one configurations of $B$,
$\gamma$ and $S$, there exists in each region of the ($x$,$y$) plane at
least one attractor and one repellor.  This is a necessary condition
for there to exist no closed orbits, and for there to be at least
one available attractor for each trajectory to end one, and one
repellor from which it can begin.  The exceptional case is $S=1$,
$-1<B<0$ and $\gamma>5/3$.  In this case there exist regions which are
either without an attractor, or without a repellor.  An example is
shown in Figure \ref{fluidplot6}.  The star in the right-hand
white region can be seen to act as an attractor, but following
trajectories in this region backwards shows that they continue to spiral outwards
forever.  Such a trajectories therefore has no simple asymptote
in the past.  Similarly, in the left-hand white region of this plot
all trajectories begin on the star, and spiral outwards forever as
they are followed forward in time.

Let us now consider further the behaviour of the scale factor, $a(t)$, as
the critical points are approached.  Trajectories approaching the
points $1$-$6$ all have evolutions similar to
those approaching the analogous points in the vacuum case.  If
condition (\ref{range}) is met, then expanding trajectories with these
points in their past all start with a big-bang singularity. Expanding
trajectories that asymptote toward these points exhibit power-law behaviour, as $t \rightarrow \infty$.  If
condition (\ref{range}) is not met, then expanding trajectories
approaching points $5$ and $6$ have a big-rip singularity in their future.

The matter dominated points, $9$ and $10$, have no analogy in the vacuum
cosmologies.  Trajectories ending on these points approach the matter
dominated solution (\ref{matter}).  When $B<1$ these points
correspond to a big bang in the past of expanding solutions, and to
power-law expansion in their future.  If $B>1$, for expanding
solutions, then there is a big rip in the future, where $a \rightarrow
\infty$ at finite $t$.

The two remaining points, $1$ and $2$, are at $y \rightarrow \pm
\infty$, and $x/y \rightarrow 0$.  In general it is required to
solve the dynamical equation (\ref{dyn3})-(\ref{constraint}), for some
particular $f(R)$, in order to find the form of $a(t)$
as these points are approached.  If $A$ and $B \rightarrow$constant in
this limit then (\ref{dyn3})-(\ref{constraint}) give
\begin{align}
w &\rightarrow w_0 (T-T_0)^{1+2 A}\\
x &\rightarrow \frac{1}{2} S (T-T_0) +\frac{1}{4}Q B w_0^2 (T-T_0)^{1+\frac{2}{B}}\\
y &\rightarrow \frac{2}{(T-T_0)}\\
z &\rightarrow z_0+4 A \ln (T-T_0),
\end{align}
where $T_0$, $w_0$ and $z_0$ are constants.  The definition of $T$,
(\ref{T}), can then be integrated to
\begin{equation}
(t-t_0) = \frac{\sqrt{3} w_0}{2 e^{\frac{z_0}{2}}} (T-T_0)^2,
\end{equation}
where $t_0$ is another constant.  The limit $y \rightarrow \infty$ can
now be seen to be reached as $T \rightarrow T_0$, and hence $t
\rightarrow t_0$.  The scale factor can then be found,
by integrating $x$, to be
\begin{align}
\ln \left( \frac{a}{a_0} \right) &= \frac{1}{4} S (T-T_0)^2 + \frac{Q
  w_0^2 B^2}{8 (1+B)} (T-T_0)^{\frac{2 (1+B)}{B}}\\
&= \frac{S e^{\frac{z_0}{2}}}{\sqrt{12} w_0} (t-t_0)+ \frac{Q
  w_0^2 B^2}{8 (1+B)} \left( \frac{2 e^{\frac{z_0}{2}}}{\sqrt{3} w_0}
  \right)^{\frac{(1+B)}{B}} (t-t_0)^{\frac{(1+B)}{B}}.
\end{align}
For $B>0$ or $B<-1$ the scale factor therefore reaches a finite value,
$a_0$, as $t \rightarrow t_0$.  The trajectory must
then be matched onto another solution at this point, in order to make
a complete history.  If $-1<B<0$ then $a \rightarrow \infty$ as these
points are approached, and there is a big rip as $t \rightarrow t_0$.

\section{The Effect of Higher Powers}

Most of the asymptotic behaviours found above correspond to
scale factors obeying power laws of the form $a \sim (t-t_0)^k$, where
$k$ is some constant.  In such cases the Ricci scalar is given by
\begin{equation}
R= \frac{6 k (2k-1)}{(t-t_0)}.
\end{equation}
Solutions approaching these asymptotes, as $t \rightarrow \pm \infty$
or $t_0$, therefore have $R \rightarrow 0$ or $\pm \infty$.  It is then straightforward to read off the values
of $A$ and $B$ for different $f(R)$.  Consider, for example, a
theory of the type
\begin{equation}
\label{powerfR}
f(R)= \sum_i c_i R^i,
\end{equation}
where the $c_i$ are constants, and the sum is over finite $i$.  In this case the functions $A$ and $B$
are given by
\begin{align}
A &= \frac{\sum_i i c_i R^{i-1}}{2 \sum_j j (j-1) c_j R^{j-1}}\\
B &= \frac{\sum_i (i-1) c_i R^i}{\sum_j j c_j R^j}.
\end{align}
It can then be seen that as $R \rightarrow 0$, $A$ and $B$ approach the constant values
\begin{equation}
A \rightarrow \frac{1}{2 (i_- -1)} \qquad \qquad \text{and} \qquad
\qquad B \rightarrow \frac{(i_- -1)}{i_-},
\end{equation}
where $i_-$ is the lowest power in $f(R)$.  Similarly, as $R
\rightarrow \pm \infty$ we have
\begin{equation}
A \rightarrow \frac{1}{2 (i_+ -1)} \qquad \qquad \text{and} \qquad
\qquad B \rightarrow \frac{(i_+ -1)}{i_+},
\end{equation}
where $i_+$ is the highest power in $f(R)$.  In such limits these
theories behave as if $f(R)=R^n$, with the
lowest power in (\ref{powerfR}) dominating as $R \rightarrow 0$, and
the highest power dominating as $R \rightarrow \pm \infty$.  Even in
the absence of a highest power in $f(R)$ (that is, for an infinite
power series) $A$ and $B$ can still approach
constant values, a necessary condition to allow the existence of the vacuum dominated solutions,
(\ref{vacuum}), and matter dominated solutions, (\ref{matter}).

The behaviour of $R$ at late times, and hence the possible asymptotic behaviour
of $a(t)$, can now be determined from the behaviour of $R$ in that
limit.  If the vacuum dominated evolution (\ref{vacuum}) is
approached, as often is the case, then this behaviour can be read off
from condition (\ref{range}).  For $1/2<B<1$ we
have at late times that $t \rightarrow t_0$, which corresponds to $R
\rightarrow \pm \infty$ and, therefore, the asymptotic behaviour of
$a(t)$ being dominated by the highest power of $R$ in the Lagrangian.
This range of $B$ corresponds to a power $R^n$ with $n>2$.  
Conversely, for $B>1$ or $<-1$
we have that $t \rightarrow \infty$ at late times, so that $R \rightarrow 0$, and
there can exist an asymptote in which $a(t)$ is dominated by the lowest power of
$R$.  These ranges of $B$ correspond to $R^n$ with $n<1/2$.
If a theory contains a low power of $R$ in its Lagrangian there can
then exist a vacuum dominated
asymptote where $R \rightarrow 0$, and in which the lowest power of $R$
dominates. However, if that same theory contains high powers of $R$ in
its Lagrangian there can also exist an asymptote
where $R\rightarrow \infty$, and in which the highest power of $R$ dominates.
Theories with multiple powers of $R$ can therefore have their
late-time asymptotics dominated by either the lowest
\textit{or} highest power of $R$.

To illustrate these points further will now consider some example theories.

\subsection{Adding Higher Powers to General Relativity}

Consider adding to the Einstein-Hilbert action higher powers of $R$,
such as $R^2$ and $R^3$.  Such modifications have often been
considered as possible UV `corrections' to gravity, motivated by
attempts to construct a perturbatively re-normalisable quantum field
theory of gravity.  It is frequently assumed that these higher-order
`corrections' to gravity should dominate the earliest stages of the
Universe's evolution.  Let us investigate this possibility in the
frame-work that has been constructed above.

\begin{figure}[htb]
\begin{center}
\epsfig{figure=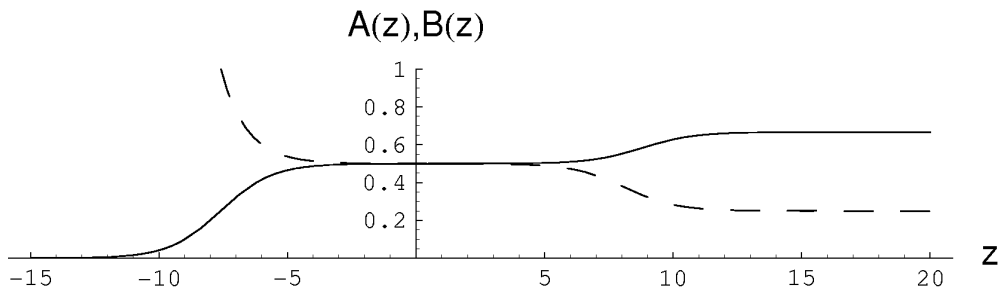,height=4cm}
\caption{The functions $A(z)$ and $B(z)$ for the theory $f(R)=R+10^3
  R^2+0.1 R^3$.  The dashed line corresponds to $A$, and the solid
  line to $B$.}
\label{AB1}
\end{center}
\end{figure}

Figure \ref{AB1} shows the functions $A(z)$ and $B(z)$ when
$f(R)=R+10^3 R^2+0.1 R^3$.  The dashed line in this plot is $A$, and
the solid line is $B$.  One can see the linear term
dominates at low $R$, the quadratic term at intermediate
$R$, and the cubic term at high $R$. 

Consider a trajectory in the `high-curvature' regime, $z \gtrsim 12$,
so that $B \simeq 2/3$.  In this case the stability of the various
critical points, and their locations, are shown in Figure
\ref{fluidplot1}.  The only expanding attractor is the vacuum
dominated solution, (\ref{vacuum}), which in this case corresponds to
$a \sim (t-t_0)^{-10}$.  As this attractor is approached we have $z^{\prime} >0$, so trajectories
are pushed further into the the high $R$
regime, and toward an eventual big rip.  For a trajectory at
high-curvature to find its way down into the intermediate curvature
regime, where $R^2$ dominates, it must therefore be either collapsing,
or possibly in the early transient stages before an asymptotic attractor is
approached.

Now consider a trajectory in the $R^2$ dominated regime, $-3 \lesssim
z \lesssim 5$, where $B \simeq 1/2$.  The ($x$,$y$) plane looks very similar to Figure
\ref{fluidplot1} in this case, but with the vacuum dominated points $5$
and $6$ displaced a little.  Again, the only
expanding attractor is the vacuum dominated solution, which in this
case corresponds to exponential expansion, (\ref{exp}).
This point is then semi-stable, in the sense described in Appendix
\ref{appendix}.  Here we have $A B_{,z} >0$, so the vacuum dominated
exponential expansion will eventually end with the point becoming a
saddle in the phase space.  Trajectories can then move to the higher
or lower $R$ regimes.

The picture of a universe starting off expanding with the highest
power in $f(R)$ dominating, and moving successively down through lower
powers of $R$ until $f \sim R$ is reached does not appear to be a generic
situation.  In fact, if a higher power of $R$ (other than $R^2$) dominates the
gravitational dynamics, then it appears that expanding trajectories are
generically forced to higher $R$, rather than lower.  The $R^2$ case
is exceptional, and in this case expanding
vacuum dominated trajectories do appear able to move to
the lower $R$ regime, although even in this case it does not seem generic.

\subsection{General Relativity as a Higher Power}

Consider now a gravitational Lagrangian with an Einstein-Hilbert term,
and both higher and lower powers of $R$.  Lower powers of $R$ have
been of interest recently, as they can correspond to
cosmologies that accelerate at late times.

\begin{figure}[htb]
\begin{center}
\epsfig{figure=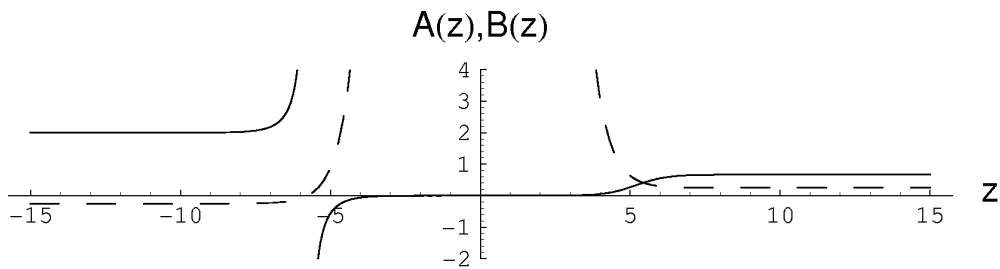,height=4cm}
\caption{The functions $A(z)$ and $B(z)$ for the theory
  $f(R)=R+10^{-5} /R +10^{-5} R^3$.  The dashed line corresponds to $A$, and the solid
  line to $B$.}
\label{AB2}
\end{center}
\end{figure}

In Figure \ref{AB2} we show the form of $A(z)$ and $B(z)$ for the
theory $f(R)=R+10^{-5} /R +10^{-5} R^3$.  At high $R$ the $R^3$ term
dominates, at intermediate $R$ the Einstein-Hilbert term dominates,
and at low $R$ the inverse power, $1/R$, dominates.
As before, when $R^3$ dominates the only expanding attractor at late times
corresponds to increasing $R$.  So again, only contracting
solutions, or trajectories in the early transient stages, can make it
down to the lower curvature regimes.  A solution in the intermediate
curvature regime, $-3 \lesssim z \lesssim 3$, is dominated here by the
Einstein-Hilbert term, with $B\simeq 0$.  In this case the dynamical equations
reduce to the usual Friedmann ones, and so $R$ decreases for
expanding solutions, allowing for a transition to the $1/R$ regime, $z
\lesssim -12$, where $B \simeq 2$.

The stability properties and location of critical points in the
$1/R$ regime are shown in Figure \ref{fluidplot2}.  In this situation
we again have that the only expanding attractor is the vacuum
dominated solution, (\ref{vacuum}).  In this case, however, $A<0$, so
this attractor corresponds to decreasing $R$.  Trajectories
asymptoting toward this point are then pushed further and further
into the $1/R$ dominated regime.  If we were to have included an even
lower power of $R$, say $1/R^2$, then expanding solutions heading
toward the attractor solution would be pushed toward the regime in
which this power dominates.

Expanding solutions in regimes dominated by the Einstein-Hilbert term,
or lower powers of $R$, generically appear to move to lower and
lower $R$ as they continue to expand.  The term $R^2$ then
acts as a watershed: expanding universes dominated by higher powers of $R$
appear to evolve toward higher $R$, while expanding universes dominated by
lower powers move to lower $R$.  For the case of high $R$
this corresponds to divergent expansion toward a big rip, and for
low $R$ it corresponds to (possibly accelerating) eternal expansion.

\subsection{Infinite Power Series}

\begin{figure}[htb]
\begin{center}
\epsfig{figure=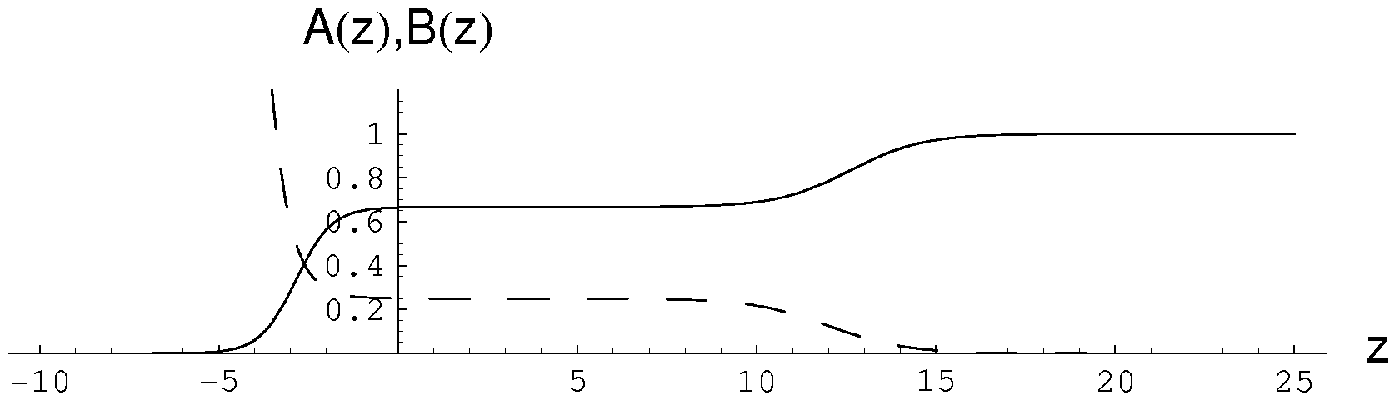,height=4cm}
\caption{The functions $A(z)$ and $B(z)$ for the theory
  $f(R)=(R+100 R^3) e^{ 10^{-5} R }$.  The dashed line corresponds to $A$, and the solid
  line to $B$.}
\label{AB3}
\end{center}
\end{figure}

Having considered the case of single higher or lower powers dominating the
gravitational Lagrangian, let us now
consider infinite power series.  There are, of course, very
many functional forms that one may choose for $f(R)$ to illustrate the case of an
infinite power series.  Here we consider
$f(R)=(R+100 R^3) e^{ 10^{-5} R }$.  The corresponding
functions $A(z)$ and $B(z)$ are shown in Figure \ref{AB3}.  It can be
seen that there are three different regimes: low $R$ at $z \lesssim
-5$, where the Einstein-Hilbert term dominates with $B \simeq 0$,
intermediate $R$ at $0 \lesssim z \lesssim 10$, where $R^3$ dominates
with $B\simeq 2/3$, and high $R$ at $z \gtrsim 15$, where the
exponential dominates with $B \simeq 1$.

The high $R$ regime now corresponds to $B\simeq 1$, which is one of the
special cases in Appendix \ref{appendix}.  The vacuum dominated
point $5$ is again the only expanding attractor, and now this point
corresponds to exponential growth of the scale factor, as in
(\ref{exp}).  This point is then stable, as $A_{z} \leqslant 0$, but
with $z^{\prime}=0$.  While it can be the case that
single higher powers dominate at intermediate regimes ($R^3$ in the region $0
\lesssim z \lesssim 10$, here), the solutions heading toward
the expanding attractor in such regimes are forced to higher $R$.  The
highest $R$ regime here, however, is at $B \simeq 1$.  The late-time attractor
then no longer corresponds to a big rip, but to exponential expansion,
with $R \rightarrow$constant, instead of $\infty$.  The
behaviour of infinite power series in $R$ is therefore qualitatively
different to the finite series considered previously.  As
before, only collapsing solutions, or possibly those in the early transient
stages of their evolution, appear able to make it down to the low
curvature regime.

\section{Discussion}

We have considered in this paper the evolution of spatially flat FRW
universes governed by $f(R)$ theories of gravity.  The
Friedmann equations (\ref{Friedmann1})-(\ref{conservation}) were transformed into an autonomous system of
first-order differential equations, and a dynamical systems analysis
was performed.  The location and stability of all critical points in
the phase space were found, for both vacuum
and perfect fluid cosmologies, and for general $f(R)$.  It was shown
that the simple power-law solutions, given by equations (\ref{vacuum})-(\ref{radiation}), often
act as the early and late-time asymptotes of the general solution.

The general behaviour of $f(R)$ FRW
cosmologies is complicated.  The phase space of
solutions is often divided into sub-spaces by invariant manifolds,
through which the trajectories describing the general solution cannot
pass.  The asymptotic past of general solutions can contain points of
inflexion, or big-bang singularities that can be approached in different ways.  Similarly, future behaviour
can be seen to be able to asymptote toward matter dominated expansion,
(\ref{matter}), vacuum domination, (\ref{vacuum}), or various
other forms.  Whatsmore,
the three dimensional phase space of solutions, in the presence of a
perfect fluid, allows for the possible existence of strange attractors, and
chaotic behaviour.

Nevertheless, despite the complicated behaviour exhibited by the
general solutions, it is still possible to make statements about the
effects of modifying the Einstein-Hilbert action to more general
functions of $R$.  It can be said that theories that
contain lower powers of $R$ often have a stable
asymptote that corresponds to the expanding vacuum dominated solution,
(\ref{vacuum}), of that lowest power, and that therefore behave as if
governed by a gravitational Lagrangian of the form (\ref{low}) at late
times.  However,
if a theory contains any higher powers of $R$ then there is
also often a stable asymptote that corresponds to the expanding vacuum dominated
solution, (\ref{vacuum}), of that highest power, and that therefore
behaves as if governed by a Lagrangian of the form (\ref{high}) at
late times.
Theories containing both higher and lower powers of $R$ can then
asymptote, at late times, to regimes in which either the lowest {\it
  or} highest powers or $R$ dominate.  In either case the
consequent evolution is that of a gravitational Lagrangian dominated
by a single power of $R$.

Which solutions asymptote to high $R$ domination, and which to low
$R$ domination, depends on the initial conditions, and the form of
$f(R)$.  Using illustrative examples we have shown that if a power of
$R$ higher than $R^2$ dominates at some point
then the generic behaviour of expanding solutions is to higher $R$.  If
a power of $R$ lower than $R^2$ dominates, then the trend is to lower
$R$.  A term $R^2$ in the Lagrangian is then a special case;
if it dominates then the expanding attractor corresponds to
exponential growth, and acts as a separatrix between the higher
or lower powers of $R$ dominating the future dynamics of the Universe.
Those solutions with low $R$ dominating at late-times expand
eternally, while those with high $R$ dominating approach either a big-rip
singularity, or exponential expansion.  Big-rips often occur if
there exists a single power of $R$ that dominates at late times, and exponential expansion occurs if $B \rightarrow
1$, as is the case for some infinite power series, such as $f \sim \exp \{
R \}$.  This late-time acceleration does not appear to be a good
candidate for the apparent acceleration we observe, however, as it cannot follow
from a period of Einstein-Hilbert domination in which $R \rightarrow 0$.

The asymptotic past of the general solutions is similarly
complicated.  A variety of behaviours seems possible, including
big-bang singularities and bounces, where the scale factor reaches a
non-zero minimum.  Big-bang singularities are
often approached with the scale-factor behaving as in the Tolman
solution, (\ref{radiation}).  It is interesting to note that even
trajectories which undergo an early period of inflation (such as those
solutions in which a power of $R^2$ dominates at some point) do not
generically follow such expansion indefinitely into the past, but
rather have a big bang or bounce at some point in their past, prior to
the onset of inflation.

While the behaviour found here is quite complicated, with
numerous different asymptotes possible, it is certainly
not the most general case that one may consider.  We have limited
ourselves here to spatially flat FRW universes.  Relinquishing the
criterion of spatial flatness would lead to more complicated
behaviours still, and one may also consider inhomogeneous and/or
anisotropic cosmologies, or cosmologies with multiple fluids.  It is
not clear whether or not the behaviour identified above would hold in
these more general cases or not.  What does seem clear, however, is that
the simple picture of the highest powers of $R$ dominating at early times,
and lower powers dominating at late times, is unlikely to be accurate.

\appendix
\section{Stability of `Vacuum Dominated' Solutions when $\mathbf{B=1/2}$
or $\mathbf{1}$}
\label{appendix}

When $B=1/2$ or $1$ the vacuum dominated solutions take an exponential
form $a \sim e^{c(t-t_0)}$, instead of the usual power-law form,
(\ref{vacuum}).  In this case the parameter $z=\ln \vert R \vert$
does not diverge to $\pm \infty$, and so the stability analysis must be
modified from the power-law case.  We will investigate the stability
of these solutions here.

\subsection{Vacuum Cosmologies}

In the vacuum cosmologies, considered in section 3, points $5$ and $6$
correspond to exponential expansion when $B \rightarrow 1/2$ or
$1$.  Consider first the case $B \rightarrow 1/2$.   Perturbing $\theta \rightarrow \theta + \delta \theta$
and $z \rightarrow z + \delta z$ we then have that the evolution
equations (\ref{theta}) and (\ref{R}) are, to linear order,
\begin{align}
\delta \theta^{\prime} &= \mp \frac{\sqrt{3 Q}}{2} \delta \theta \pm
\frac{4 \sqrt{Q}}{\sqrt{3}} B_{,z} \delta z\\
\delta z^{\prime} &= \pm A \sqrt{\frac{Q}{3}} \delta \theta,
\end{align}
where the upper branch corresponds to point $5$, and the lower branch to
point $6$.  It can now be seen that the eigenvalues to the equations $\delta
z^{\prime} = \lambda_i \delta z$ and $\delta \theta^{\prime} =
\lambda_i \delta \theta$ are given by the two roots of
\begin{equation}
\lambda_i^2 \pm \frac{\sqrt{3 Q}}{2} \lambda_i - \frac{4}{3} A B_{,z}
Q =0.
\end{equation}
On recognising that these solutions have $Q=1$, it can
be seen that if $A B_{,z}>0$ then both points are
saddles.  Alternatively, if  $A B_{,z} <0$ then point $5$ is stable,
while point $6$ is unstable.  We will also be interested in the case
when $B_{,z}\simeq 0$.  Point $5$ is then stable in $\theta$,
while $\delta z^{\prime}=0$.  Small fluctuations (as one would expect
to occur in a thermal de Sitter space such as this) will then
gradually shift the value of $z$.  Unless the theory has $B_{,z}=0$
for all $z$, there will then come a time at which $B_{,z} \neq 0$, after
which the point becomes an attractor proper, or a saddle, depending on
the sign of $A B_{,z}$.  We will refer to this situation of temporary
stability, followed by saddle behaviour, as `semi-stable'.

Now consider the case $B \rightarrow 1$.  In this case it is instructive to use
the definitions of $A$ and $B$ to see that
\begin{equation}
\label{AB}
A=\frac{(1-B)}{2 (B+B_{,z})}.
\end{equation}
For $B_{,z} \neq -1$ we then have that $B \rightarrow 1$ corresponds to
$A \rightarrow 0$.
Perturbing $\theta$ and $z$, as before, we then find that the
linearised evolution equations become
\begin{align}
\delta \theta^{\prime} &= \mp 2 \sqrt{\frac{Q}{3}} \delta \theta \pm
\frac{\sqrt{3 Q}}{2} B_{,z} \delta z\\
\delta z^{\prime} &= \sqrt{\frac{Q}{3}} A_{,z} \delta z,
\end{align}
so that the relevant eigenvalues are given as the roots of
\begin{equation}
\lambda_i^2 \pm \sqrt{\frac{Q}{3}} (2-A_{,z}) \lambda_i -
\frac{2}{3} A_{,z} Q=0.
\end{equation}
Again, $Q=1$, so the condition for points $5$ and $6$ to be saddles is
now $A_{,z}>0$.  If $A_{,z}<0$ then point $5$ is an attractor, and
point $6$ a repellor.  Now, if $A_{,z}=0$, then point $5$ can be
semi-stable, depending on the sign of $A_{z}$.

\subsection{Perfect Fluid Cosmologies}

In the perfect fluid cosmologies we also find that the vacuum
dominated solution corresponds to exponential expansion when $B
\rightarrow 1/2$ or $1$.  Consider first the case $B \rightarrow 1/2$.  In this case a perturbative
expansion in $\phi$ and $z$ gives the evolution equations
(\ref{dyn4}), (\ref{rprime}) and (\ref{phi2}), to linear order, as
\begin{align}
\delta \theta^{\prime} &= \pm \frac{(8 B_{,z} \delta z-3 \delta
  \theta)}{(1-r)}\\
\delta z^{\prime} &= \pm \frac{(2 A \delta \theta)}{(1-r)}\\
r^{\prime} &= \pm \frac{3 \gamma}{2},
\end{align}
where the upper sign is for point $5$ and the lower sign for point
$6$.  We are now looking for eigenvalues, $\lambda_i$, such that
\begin{equation}
\label{eig}
\delta \theta^{\prime} = \frac{\lambda_i}{(1-r)} \delta \theta \qquad
\qquad \text{and} \qquad \qquad \delta z^{\prime} =
\frac{\lambda_i}{(1-r)} \delta z.
\end{equation}
If both of these eigenvalues are negative, and $r^{\prime}>0$, then the
point is stable.  If they are both positive, and $r^{\prime}<0$, then
the point is unstable.  Any other combination we will call a saddle
point.  The values of $\lambda_i$ can be seen to be given by the roots
of
\begin{equation}
\lambda_i^2 \pm 3 \lambda_i -16 A B_{,z} =0.
\end{equation}
It can then be seen that for $A B_{,z} >0$ both points are saddles,
while for $A B_{,z} <0$ point $5$ is stable and point $6$ is
unstable.  $A B_{,z}=0$ again corresponds to point $5$ being semi-stable.
The stability properties of points $5$ and $6$ in the
presence of a perfect fluid when $B \rightarrow 1/2$ are therefore identical to the
corresponding points in the vacuum cosmologies, considered above.

Finally, consider the case $B \rightarrow 1$.  Equation (\ref{AB}) shows
that in this limit we again have $A \rightarrow 0$.  The perturbed evolution
equations are then given as
\begin{align}
\delta \theta^{\prime} &= \pm \frac{(3 B_{,z} \delta z-4 \delta
  \theta)}{\sqrt{2} (1-r)}\\
\delta z^{\prime} &= \pm \frac{\sqrt{2} A_{,z} \delta z}{(1-r)}\\
r^{\prime} &= \pm \frac{\sqrt{2}}{4} (1+3 \gamma),
\end{align}
where upper signs correspond to point $5$, and lower signs to
point $6$.  The eigenvalues, $\lambda_i$, are now
given by the roots of
\begin{equation}
\lambda_i^2 \pm \sqrt{2} (2-A_{,z}) \lambda_i - 4 A_{,z}=0.
\end{equation}
For $A_{,z}>0$ both of points $5$ and $6$ are 
saddles, while for $A_{,z}<0$ point $5$ is stable and point $6$ is
unstable.  $A_{,z}=0$ gives point $5$ as semi-stable.  Again, the
stability properties of these points is identical to the
analogous points in the vacuum cosmologies.

\bigskip

\bigskip

\leftline{\bf \Large{Acknowledgements}}

I would like to thank John Barrow for suggestions, and to acknowledge the support of Jesus College, Oxford.

\end{document}